\documentclass[sigconf, dvipsnames]{acmart}
\usepackage[acronym]{glossaries}
\usepackage{amsmath,amsfonts, amsthm,mathtools}
\usepackage{graphicx}
\usepackage{textcomp}
\usepackage{colortbl}
\usepackage{flushend} 
\usepackage{float}
\usepackage{subfigure}
\usepackage{multirow}
\usepackage{multicol}
\usepackage{xspace}
\usepackage{comment}
\usepackage{soul}
\usepackage{multirow}
\usepackage{stfloats} 
\usepackage{url}
\usepackage{longtable,booktabs}
\usepackage[version=4]{mhchem}
\usepackage[compact]{titlesec}
\usepackage{setspace}
\usepackage{mathrsfs}
\usepackage{annotate-equations}
\usepackage{mathrsfs}
\usepackage{algorithm}
\usepackage{algpseudocode}
\def\secformat{\bfseries}
\usepackage{listings}
\usepackage[most]{tcolorbox}
\newtcolorbox[auto counter]{finding}[1][]{%
    colback=blue!5,           
    colframe=blue!40,         
    boxrule=0pt,              
    leftrule=2mm,             
    sharp corners,            
    before upper={\textbf{Finding~\thetcbcounter:}~}, 
    fontupper=\normalfont,    
}

\definecolor{neongreen}{rgb}{0.0, 1.0, 0.0} 
\definecolor{neonpink}{rgb}{1.0, 0.07, 0.58} 
\sethlcolor{neonpink}

\usepackage{enumitem}

\definecolor{secblue}{HTML}{00008B}
\definecolor{subsecblue}{HTML}{0000FF}
\definecolor{subsubsecblue}{HTML}{0047AB}
\definecolor{captionblue}{HTML}{0000FF}
\definecolor{tableheader}{HTML}{203864}
\definecolor{paragraph}{HTML}{3B5E7F}
\definecolor{pgtext}{HTML}{0000CD}
\definecolor{hdrgray}{HTML}{BFCDDB}%
\definecolor{ltgray}{HTML}{DCDCDC}%
\definecolor{cellcolor}{HTML}{BFCDDB}

\titleformat{\subsection}{\secformat\color{subsecblue}}{\thesubsection}{0.5em}{}
\titleformat{\subsubsection}{\secformat\color{subsubsecblue}}{\thesubsubsection}{0.5em}{}

\glsdisablehyper
\makeglossaries   
\newacronym{tl}{TL}{Transfer Learning}
\newacronym{hpc}{HPC}{High Performance Computing}
\newacronym{ml}{ML}{Machine Learning}
\newacronym{nlp}{NLP}{Natural Language Processing}
\newacronym{amm}{AMM}{Abstract Machine Model}
\newacronym{sipt}{M\textsubscript{$\mathcal{SA}$}}{\texttt{Stacked Input Alignment Model}}
\newacronym{ipt}{M\textsubscript{$\mathcal{A}$}}{\texttt{Input Alignment Model}}

\newacronym{cn}{$\mathcal{M}odel\mathcal{X}$}{\texttt{Cross Prediction Model}}
\newacronym{mse}{MSE}{Mean Square Error}
\newacronym{nn}{NN}{Neural Network}
\newacronym{lp}{LP}{Linear Probing}
\newacronym{ft}{FT}{Fine-tuning}
\newacronym{ind}{IND}{In-Distribution}
\newacronym{ood}{OOD}{Out-of-Distribution}
\newacronym{fsl}{FSL}{Few-Shot Learning}
\newacronym{sssp}{SSSP}{Single-Source Shortest Path}
\newacronym{ae}{AE}{Auto Encoder}
\newacronym{vae}{VAE}{Variational Auto Encoder}
\newacronym{tab}{TabNet}{Transformer}
\newacronym{llm}{LLM}{Large Language Model}
\newacronym{llms}{LLMs}{Large Language Models}

\definecolor{sg}{HTML}{45B39D}

\definecolor{sgl}{HTML}{A2D9CE}
\definecolor{bl}{HTML}{5499C7}
\definecolor{bll}{HTML}{A9CCE3}
\definecolor{or}{HTML}{D35400}
\definecolor{orl}{HTML}{EDBB99}
\definecolor{usecase}{HTML}{008FAC}
\definecolor{deepskyblue}{HTML}{00BFFF}
\definecolor{purple}{HTML}{C4A9FF}
\definecolor{bb}{HTML}{DAE3F3}
\definecolor{oo}{HTML}{FBE5D6}
\definecolor{gry}{HTML}{EDEDED}
\definecolor{yl}{HTML}{FFC000}
\definecolor{yll}{HTML}{FFFFE0}

\usepackage{tikz}

\definecolor{green}{rgb}{0.1,0.1,0.1}

\newcommand{\sysname}{\textsc{Wander}\xspace}

\newcommand{\tzi}[1]{\textcolor{red}{FIX: #1}\xspace}

\usepackage{diagbox}
\usepackage{fontawesome5}

\usepackage{caption}
\usepackage{dcolumn, array, booktabs}
\newcolumntype{.}{D{.}{.}{-1}}

\AtBeginDocument{%
  }

\setcopyright{acmlicensed}
\copyrightyear{2018}
\acmYear{2018}
\acmDOI{XXXXXXX.XXXXXXX}
\acmConference[Conference acronym 'XX]{Make sure to enter the correct
  conference title from your rights confirmation email}{June 03--05,
  2018}{Woodstock, NY}
\acmISBN{978-1-4503-XXXX-X/2018/06}

\begin{document}

\title{\sysname: An Explainable Decision-Support Framework 
for HPC}


\author{Ankur Lahiry}
\affiliation{%
  \institution{Texas State University}
  \city{San Marcos}
  \country{United States}}
\email{vty8@txstate.edu}

\author{Banooqa Banday}
\affiliation{%
  \institution{Texas State University}
  \city{San Marcos}
  \country{United States}}
\email{banooqa@txstate.edu}

\author{Yugesh Bhattarai}
\affiliation{%
  \institution{Texas State University}
  \city{San Marcos}
  \country{United States}}
\email{yugesh@txstate.edu}

\author{Tanzima Z. Islam}
\affiliation{%
  \institution{Texas State University}
  \city{San Marcos}
  \country{United States}}
\email{tanzima@txstate.edu}








\begin{abstract}
High-performance computing (HPC) systems expose many interdependent configuration knobs that impact runtime, resource usage, power, and variability. Existing predictive tools model these outcomes, but do not support structured exploration, explanation, or guided reconfiguration. We present \sysname, a decision-support framework that synthesizes alternate configurations using counterfactual analysis aligned with user goals and constraints. We introduce a composite trade-off score that ranks suggestions based on prediction uncertainty, consistency between feature-target relationships using causal models, and similarity between feature distributions against historical data. To our knowledge,~\sysname is the first such system to unify prediction, exploration, and explanation for HPC tuning under a common query interface. Across multiple datasets, \sysname generates interpretable and trustworthy, human-readable alternatives that guide users to achieve their performance objectives.


\end{abstract}

\ccsdesc[500]{Do Not Use This Code~Generate the Correct Terms for Your Paper}
\ccsdesc[300]{Do Not Use This Code~Generate the Correct Terms for Your Paper}
\ccsdesc{Do Not Use This Code~Generate the Correct Terms for Your Paper}
\ccsdesc[100]{Do Not Use This Code~Generate the Correct Terms for Your Paper}

\keywords{Counterfactual, What-if exploration, Recommendation, Causal Modeling, Generative Modeling, AI, HPC, Performance}


\maketitle

\section{Introduction}
\gls{hpc} systems are essential infrastructure for scientific advancement, powering large-scale simulations, data-intensive analyses, and modeling efforts across domains such as climate science, genomics, materials engineering, and astrophysics. These systems enable computations at scales that are otherwise infeasible due to the complexity and size of modern scientific workloads. However, effective utilization of~\gls{hpc} systems remains a challenge. As architectures become increasingly heterogeneous—with combinations of CPUs, GPUs, and specialized accelerators—operational decision-making becomes more complex.

Performance optimization now spans a high-dimensional configuration space, involving runtime parameters such as hardware selection, resource allocation, execution strategy, and power constraints.
The impact of these decisions is nontrivial. Prior work has shown that a configuration that yields optimal performance in one scenario can cause up to $4\times$ variability under slightly different conditions~\cite{patki2019performance}. This makes "performance" not a single point objective, but a non-convex, multi-objective trade-off space, whose characteristics must be understood to make informed, robust decisions.

We introduce~\sysname--\textit{What-if Analysis and Navigation for Decision-support with Explainable Reasoning}--a framework that guides various \gls{hpc} stakeholders including users and software systems through complex performance trade-offs by illuminating actionable configurations.

This paper focuses on operational decision-making: how to configure runtime parameters such that their interaction with an application's performance profile--modeled by predictive models--yields satisficing performance. 
A satisficing configuration is defined as any point within a user-defined region of acceptable performance based on user defined performance objectives such as runtime, power, or reliability thresholds. 

In the case of recommendations, the system prioritizes satisficing solutions—configurations that meet user-defined thresholds across multiple objectives without necessarily being optimal. However, the framework itself is not limited to satisficing; if users specify a particular trade-off or counterfactual target, \sysname simply returns the appropriate configuration. It is only when users request the ``best" configuration under multiple objectives that the problem is explicitly framed as a satisficing query.

Despite extensive literature in performance modeling, scheduling, and autotuning, there remains a gap in tools that enable end-to-end what-if exploration by both users and software systems. In the absence of such tools, users are often forced to repeatedly build ad hoc performance models, conduct analysis over specific datasets, and manually interpret outcomes—all of which require domain expertise and are difficult to generalize. This lack of systematized what-if reasoning leads to brittle, non-portable workflows that do not scale or adapt across workloads and platforms. Importantly, most existing tools fail to link observational performance data with \textbf{interventional decision support}--enabling users to simulate outcomes of changes before enacting them.

To address this gap, we introduce \sysname, a decision-support tool that unifies predictive performance modeling, generative counterfactual reasoning, and query-driven recommendation within a single extensible framework. Unlike black-box predictors, \sysname is designed to simulate, explain, and quantify the outcomes of configuration changes by modeling the underlying causal structure of the system. The framework allows users to ask diverse what-if questions across three core decision services:
(1) Recommendation: ``What runtime configuration can meet a target performance goal?"
(2) Prediction: ``What performance can I expect from a given configuration?"
(3) Counterfactual reasoning: ``What minimal changes would yield an improved performance under a constraint?"

To illustrate the capabilities of~\sysname, we use an example focused on a scientific simulation running on a heterogeneous cluster. A user investigates how increasing the number of compute nodes \texttt{num\_node} from a base configuration to a certain number might influence \texttt{node\_power\_consumption}. The objective of this investigation is to estimate whether scaling up resources can reduce overall runtime without exceeding the predefined energy or power constraints. This example threads through the paper, showcasing how modeling, counterfactual reasoning, and recommendation converge to produce actionable decisions.


\sysname is designed to support flexible system integration. It can serve end-users directly, helping scientists and engineers reason about how to run their jobs. At the same time, the API layer can be detached and integrated into job schedulers or workflow orchestrators, enabling automated decision support in larger \gls{hpc} pipelines. Although the current implementation focuses on individual job-level decisions, the architecture can be extended to reason about site-wide or even cross-site optimization strategies. However, we note that we currently lack datasets to explore the latter scenarios.

To increase trust , interpretability and accountability, we propose a novel evaluation methodology using structural causal models (SCMs) that assess the robustness and internal consistency of generated counterfactuals. In addition, we implement ensemble-based uncertainty quantification, allowing the system to present confidence intervals for each recommended configuration. Furthermore we incorporate outlier detection to find out how our method is prone to the accountability. Together, these mechanisms enable risk-aware, transparent decision-making for both users and automated systems.

Instead of targeting specific user roles, \sysname organizes functionality around three query types that cut across stakeholder categories: (1) Recommendation queries support users seeking feasible runtime configurations to meet performance thresholds.
(2) Prediction queries allow system administrators to simulate the impact of policy changes on center-wide throughput or power efficiency.
(3) Counterfactual queries assist stakeholders in reasoning about alternate configurations to improve past performance outcomes.

Our tool is designed to support these diverse scenarios:
(1) Application scientists can explore how to run their job, what type of configurations they should use to achieve certain threshold (recommendation), (2) System administrators and center managers can explore site-wide policies and predict their impact on the overall throughput of their \gls{hpc} facilities (explore and predict), (3) stakeholders can explore ``what-if" scenarios by asking the tool about how to change their previous decisions to achieve a performance objective (counterfactual).
Performance engineers can test tuning strategies, evaluate trade-offs between energy and performance, and experiment with hypothetical deployment scenarios without risking production instability. The same tool could also be used in an educational environment for training the next-generation computational scientists by enabling them to learn about relationships across many operational parameters and how those impact performance trade-off to provide with initial idea. They do not have to be experts to start using an \gls{hpc} system efficiently.

While~\sysname provides recommendations based on historical data and causal structure, it is not a real-time scheduler, nor does it replace domain expertise. Rather, it serves as a decision-support tool that enables informed decision-making for various stakeholders by offering interpretable guidance, highlighting trade-offs, and quantifying the risks of alternative configurations. It does not prescribe a single ``best'' answer but offers multiple diverse, feasible options--each accompanied by predicted outcomes and confidence levels. Moreover, currently~\sysname is only used for operational decisions such as how to run a job. In future, the backend can be extended to also guide users for critical code-related decisions; however, there are already strong tools such as CoPilot etc., so the need for us to expand our tool for that use case is not as critical. However, there is no tool for operational decision making (that is, how to set up the runtime environment of an application, how those impact multiple performance metrics, and based on that recommending users or software systems).

\noindent The key contributions of this paper are:
\begin{itemize}
    \item \textbf{Formalize} two decision-making modes—\emph{prescriptive} and \emph{exploratory}—and map common \gls{hpc} configuration queries to these services.

    \item \textbf{Explain} how and why certain configuration changes influence performance outcomes.

    \item \textbf{Design} a novel evaluation methodology that combines structural causal modeling, predictive accuracy, causal graph alignment, and uncertainty quantification to assess the trustworthiness of decisions.

    \item \textbf{Implement} \sysname, a modular decision-support framework, which will be open-sourced.



    \item \textbf{Demonstrate} \sysname across multiple \gls{hpc} datasets, enabling varied performance decisions through a unified interface.
\end{itemize}



\section{Background}

\subsection{Conditional and Unconditional Generation}
Recommender systems use a variety of techniques to predict user preferences and deliver personalized recommendations. These include Collaborative Filtering~\cite{schafer2007collaborative}, Matrix Factorization~\cite{koren2009matrix}, and more recently, Self-supervised Learning Models~\cite{yu2023self}. While effective, these methods rely heavily on user-specific historical data, which limits their generalization in data-scarce environments. Additionally, traditional approaches such as neighborhood-based collaborative filtering become computationally expensive when applied to heterogeneous data~\cite{chen2020efficient}.

To address these challenges, generative models have emerged as a popular strategy for synthetic data generation~\cite{deldjoo2024recommendation}. By learning from available system metrics, these models can simulate realistic performance profiles, filling in missing data regions and improving generalization to unseen scenarios. Synthetic generation can take two forms: (1) Unconditional generation samples directly from the learned distribution of system configurations and performance outcomes, making it especially useful for new users with no historical data. (2) Conditional generation allows crafting synthetic samples based on prior knowledge to simulate what-if scenarios, enabling exploration of different performance trade-offs using existing datasets.

\subsection{Structural Causal Modeling}
Structural Causal Models (SCMs)~\cite{bongers2021foundations} offer a rigorous mathematical framework for modeling cause-and-effect relationships, which enables robust reasoning about what-if scenarios. The models describe the relationships through various structural equations that link variables via directed causal pathways, providing a visual representation using Directed Acyclic Graphs (DAGs). This reasoning capability is crucial for the decision-making process to generate recommendations or what-if scenarios where causal relationships are more critical than correlation. For instance, user queries to a recommendation system--for example, a user investigates how increasing the number of compute nodes \texttt{num\_node} from a base configuration to a certain number might influence \texttt{node\_power\_consumption}. While observational data might suggest a correlation between \texttt{num\_node} and \texttt{node\_power\_consumption}, it cannot conclusively confirm whether the change in node count causes the observed change in power draw. 

The observational data might show correlation between threads and execution time, but this alone can't confirm that only increasing the threads causes the speedup. In real-world HPC systems, speedup of a simulation depends on various other factors such as processor affinity and background system load. Leveraging the power of SCM, we can analyze the effect of changing the number of nodes by keeping other variables constant. This helps us in a crucial way; by visualizing the DAG generated by SCM, we can gain interpretable insights into how the other parameters such as \texttt{threads\_per\_core}, \texttt{memory\_bandwidth}, etc. can contribute to altering the target variables passively. Instead of relying on surface-level correlations only, such a structural discovery provides a transparent view of underlying mechanisms, supporting informed decision-making in \gls{hpc} configurations.

\section{Methodology}
\begin{table}[t]
\footnotesize
\centering
\caption{Query types and how they differ.}
\label{tab:query-taxonomy}
\begin{singlespacing}
\centering
\resizebox{\columnwidth}{!}{
\begin{tabular}{p{1.2cm}p{4.2cm}}
\toprule
\textbf{Query Type} & \textbf{Semantics} \\
\midrule

\textbf{Prescriptive (Recommendation)} &

User specifies performance targets and configuration constraints; no baseline configuration required. Generates configurations meeting specified targets. \\

\hline

\textbf{Exploratory (What-If)} &

User provides baseline configuration and feature modifications. Predicts how specified changes impact listed performance metrics. \\

\hline

\textbf{Exploratory (Optimization)} &

User provides baseline, desired performance targets, and fixed constraints. Finds minimal configuration adjustments that achieve specified targets. \\

\hline

\textbf{Explanation} &

User provides baseline configuration and selects target performance metric. Generates feature importance and causal influence analysis to explain model predictions. \\

\bottomrule
\end{tabular}
}
\end{singlespacing}
\end{table}
\begin{figure}
    \centering
    \includegraphics[width=\columnwidth]{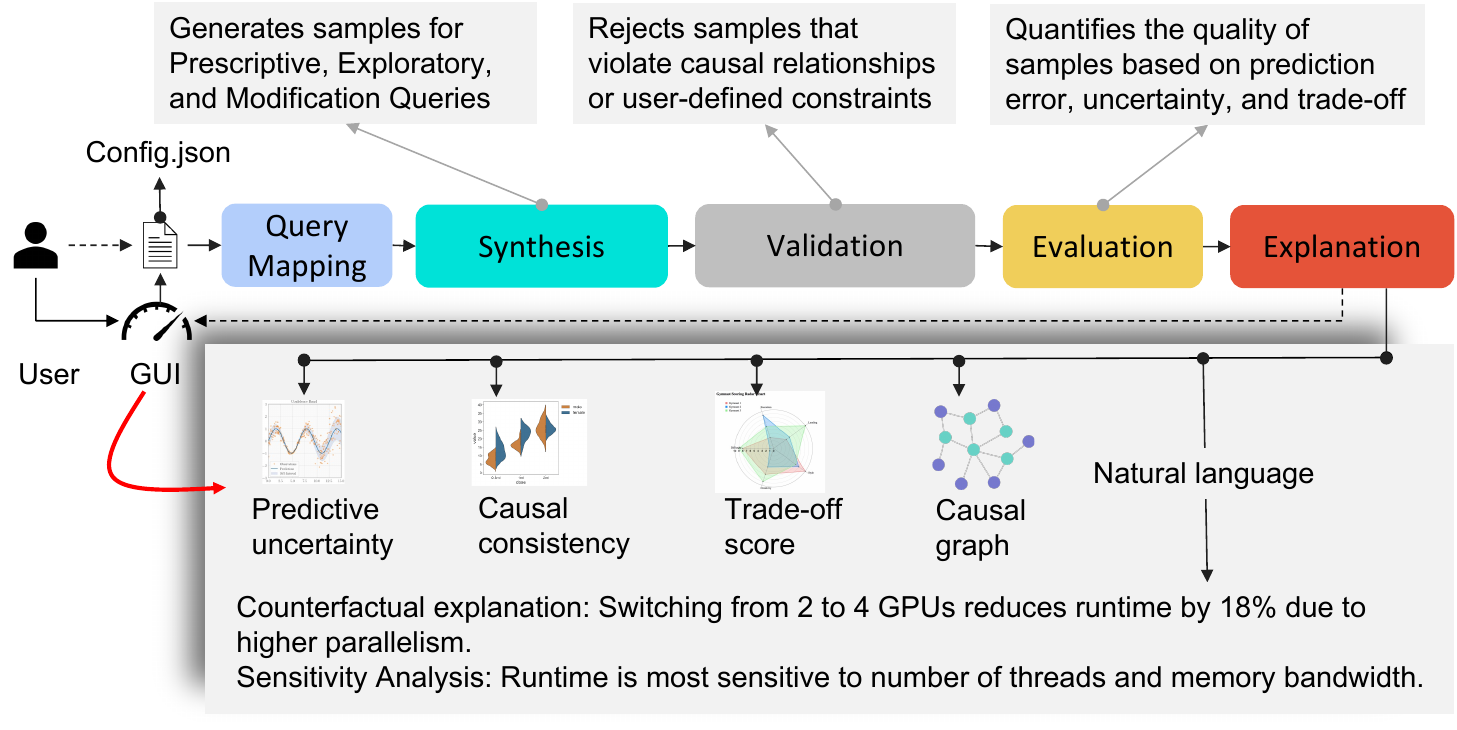}
    \caption{Overview of~\sysname. Users can interact using a \texttt{config.json} file or Streamlit GUI to specify their dataset and query. Each query is mapped to a query template in Table~\ref{tab:query-taxonomy}. Each query is transformed into a counterfactual sample generation scenario; generated samples are pruned using both constraint checking rules and causal model conformity; samples are evaluated based on various methods; the output of evaluation is summarize using natural language and presented as visualizations for users to understand.~\sysname provides a query-driven interactive decision-support environment for many~\gls{hpc} stakeholders.}
    \label{fig:overview}
\end{figure}

Before applying predictive modeling or causal analysis, \sysname performs a set of automated preprocessing steps to prepare HPC job telemetry data. 
These steps ensure numerical stability and statistical validity, particularly for downstream Structural Equation Modeling (SEM). The system automatically detects and resolves the following issues:
\vspace{-1em}
\begin{itemize}
     \item \textbf{Singular matrices:} When the covariance matrix is singular (non-invertible), Principal Component Analysis (PCA) is applied to reduce dimensionality. If singularity persists, L2 regularization is applied to stabilize model estimation.

     \item \textbf{Highly correlated features:} Features with high pairwise correlation (above a fixed threshold) are pruned to reduce multicollinearity and improve interpretability.

     \item \textbf{Zero or negative-definite covariance matrices:} A small positive constant is added to the diagonal to ensure the matrix is positive definite and invertible.
 \end{itemize}

 These preprocessing steps are implemented as modular pipelines and applied consistently across real and generated data.
\subsection{Decision Support Services using Query Mapping}

Drawing from established guidelines in human-computer interaction (HCI) and explainable AI~\cite{amershi2019guidelines, doshi2017towards, kules2008designing}, we categorize~\gls{hpc}'s decision-support needs into four categories based on the underlying intent of the query and the nature of the reasoning task. 
\begin{itemize}
    \item \textbf{Goal Oriented Decision Making} User may ask system to identify optimal performance-related goals such as minimizing runtime, reducing node power consumption, etc. (a) Single objective: decision making on a single metric; (b) Multi objective: decision making on two or more performance metrics, for example, making decision making both on runtime and node power consumption.
    \item \textbf{Exploratory What-if analysis} User's common question to simulate hypothetical scenario “What if we change \texttt{num\_nodes}, how will it affect \texttt{$node\_power\_consumption$}?” 
    
    \item \textbf{Exploratory Optimization}: User's hypothetical concern on how \% changes of a feature would 'reduce' the target metrics. For example, "what configuration changes would reduce memory stalls by 20\% while maintaing throughput?"

    \item \textbf{Root Cause Analysis}: The objective of this study to find out the root cause of a system failure. 
    
\end{itemize}

Although our framework is comprehensive and robust, this study leaves out (1b) multi-objective optimization and (4) root-cause analysis to maintain focus and depth. We plan to incorporate them in future extensions of the work. 
This taxonomy unifies system capabilities under a defined grammar, making~\sysname easily extendible in the future.~\sysname It introduces a structured query library that can be extended in the future based on user needs that cannot be supported by the current templates. 

%

\textbf{Prescriptive Queries} arise when stakeholders seek a recommendation for a configuration that satisfies one or more performance objectives under specified constraints. 
    This type of queries are useful when users know their performance goals but not how to get there, e.g., non-expert users. The response does not explain how changes affect outcomes--only what configuration meets goals.

\textbf{Exploratory Queries} enable what-if exploration. In the forward form, users change configuration features and ask how performance metrics will respond (features $\rightarrow$ targets). In the inverse form, users specify a desired performance outcome and ask what feature changes can achieve it (targets $\rightarrow$ features). These queries require an initial configuration as context. They cannot suggest new configurations without a starting point, but they allow users to reason about performance sensitivity and alternative feasible options.
    

\textbf{Modification Queries} help users adjust an existing configuration to avoid undesirable outcomes such as job failures or performance anomalies. Unlike prescriptive queries that generate configurations from scratch, or exploratory queries that explore performance trade-offs, modification queries focus on identifying the smallest change needed to flip the outcome. This type of query requires a known baseline configuration and a binary outcome label—unlike other query types. The task is modeled as a counterfactual classification problem and is particularly suited for failure diagnosis and recovery, which are not addressed by other queries. While we do not include modification queries in this study, we view them as a promising direction for future work.

\textbf{Explanation Queries} help users to understand the ``why" behind the observations produced by other queries to explain and build trust in model outputs. This includes visualizing model uncertainty, showing whether feature-target relationships in synthesized configurations match historical data, ranking configurations based on trade-offs, and translating causal graphs into natural language explanations.

\subsection{Construct queries from user input}
As Figure~\ref{fig:overview} shows, stakeholders (users or software) can interact with~\sysname via a configuration file or GUI. If users prefer GUI, their input in the form of \texttt{<Query\_Type>}, \texttt{<feature\_list>} then GUI automatically constructs the configuration file by filling in the details for the chosen query type's template.
A complete mapping is shown in Table~\ref{tab:query-taxonomy}. A full list of dataset-specific instantiations and mappings to this taxonomy is provided in Appendix~\ref{app:appendix}.

Each of these templates is generic and reusable; is implemented using a combination of predictive modeling, counterfactual generation, causal simulation, and attribution analysis. 
Following sections describe each of the methodologies in detail.


\subsection{Synthesis} 
\label{sec:counterfactual}
\sysname uses counterfactual analysis as a unified backbone to power multiple types of decision-support queries. It allows users to simulate how system behavior would change under different configuration choices. Unlike predictive models that only forecast outcomes for given inputs, counterfactuals enable the generation of new inputs that satisfy user-defined outcomes. This inversion of reasoning--from predicting effects to searching for causes--makes counterfactual analysis the precise tool for exploratory and modification (three out of four) queries. 
It fills a key gap left by standard supervised learning: while models such as XGBoost~\cite{chen2015xgboost} can forecast performance (e.g., runtime), regular predictive methods cannot suggest what inputs would lead to desired outcomes. Counterfactuals fill this gap by solving the inverse problem where they optimize input perturbations under user constraints, with guarantees of proximity, diversity, and target satisfaction. This is especially valuable for~\gls{hpc} stakeholders who need to understand, not just predict, how to improve job outcomes or avoid performance anomalies. In~\sysname, we implement counterfactual generation over a fixed predictive model $f: \mathcal{X} \rightarrow \mathcal{Y}$, trained using XGBoost. Once trained, the model weights remain fixed during counterfactual generation, following the approach of Wachter et al.~\cite{wachter2017counterfactual}. Freezing model parameters ensures that all observed changes in predicted output yield from the modified input configuration, not from model retraining or parameter change.

\subsubsection{Mathematical Formulation}
\label{sec:cf-formulation}

Counterfactual analysis requires a baseline configuration to start from, which is $\mathbf{x} \in \mathbb{R}^d$. Given a target outcome $\mathbf{y}^* \in \mathbb{R}^k$, and a pre-trained predictive model $f$,~\sysname generates a set of $N$ counterfactuals $\{\mathbf{x}'_1, \ldots, \mathbf{x}'_N\}$ by solving the constrained optimization in Equation~\ref{eqn:cf_loss}. Users can specify $N$ in the config file or GUI. By default~\sysname uses $N=20$.
\vspace{0.2in}

\begin{equation}
\footnotesize
\resizebox{\columnwidth}{!}{$
\eqnmarkbox[Green]{cf_loss}{\{\mathbf{x}'_1, \ldots, \mathbf{x}'_N\}} = \arg\min_{\{\mathbf{x}'_i\}} \left[
\sum_{i=1}^N \eqnmarkbox[WildStrawberry]{validity}{\mathcal{L}_{\text{valid}}(f(\mathbf{x}'_i), \mathbf{y}^*)} +
\lambda_1 \eqnmarkbox[YellowOrange]{proximity}{\mathcal{L}_{\text{prox}}(\mathbf{x}, \mathbf{x}'_i)} -
\lambda_2 \eqnmarkbox[RoyalBlue]{diversity}{diversity}
\right]
$}
\label{eqn:cf_loss}
\end{equation}

\annotate[yshift=0.7em]{above,right}{cf_loss}{Optimal counterfactual set}
\annotate[yshift=0.50em]{above,right}{validity}{Ensures prediction is close\\ to specified target}
\annotate[yshift=-0.2em,xshift=-0.0em]{below,left}{proximity}{Penalizes large deviations\\ from the given config}
\resizebox{\columnwidth}{!}{
\annotate[yshift=-0.3em]{below,left}{diversity}{Promotes\\ distinct solutions}
}
\vspace{0.2in}

The first term, $\mathcal{L}_{\text{valid}}(f(\mathbf{x}'_i), \mathbf{y}^*) = \frac{1}{N} \sum_{i=1}^N (f(\mathbf{x}'_i) - \mathbf{y}^*)^2$, ensures that the predicted outcome for each synthesized configuration closely matches the user-defined target $\mathbf{y}^*$. \\
The second term, $\mathcal{L}_{\text{prox}}(\mathbf{x}, \mathbf{x}') = \sum_{k=1}^{d} \frac{|\mathbf{x}_k - \mathbf{x}'_k|}{\text{MAD}_k}$, penalizes unnecessary deviations from the original configuration to preserve realism and make suggestions actionable in practice. Here, each feature difference is normalized by its Median Absolute Deviation (MAD) to account for variability across features and ensure scale-invariant comparisons.


The third term, $\text{diversity} = \text{det}(K), \quad K_{ij} = \frac{1}{1 + \|\mathbf{x}'_i - \mathbf{x}'_j\|}$, encourages variation among counterfactuals. It computes a similarity matrix \( K \), where each entry $K_{ij}$ is the pair-wise distance between counterfactual samples $x'_i$ and $x'_j$. The determinant \( \det(K) \) quantifies how diverse the generated samples are, a higher values indicate that the counterfactuals are spread out.

\subsection{Validation}
\label{sec:validation}
We use two different types of validation methodology for rigorously ensuring that the synthesized counterfactuals are plausible, not only statistically, but also causally. Specifically, we use rule-based rejection sampling and causal modeling.

\subsubsection{Rule-Based Rejection Sampling}

Though the counterfactual generation pipeline itself can handle all the feature and target constraints correctly, it often overlooks domain-specific constraints inherent to high-performance computing (HPC) environments. The generation pipeline uses algorithm perturbations which don't account for the underlying operational or physical feasibility. For example, our pipeline is responsible for generating a configuration which projects a lower memory allocation with a higher number of GPUs allocated. Though this is a valid hypothetical configuration as it meets the target constraints correctly, we can't define this configuration as "feasible" as it conflicts with High Performance Computing policy. To improve the trustworthiness and accountability of the pipeline, we introduce a novel LLM-based approach to incorporate domain knowledge without hard-coding expert rules.

We have employed OpenAI's GPT-4o a state-of-the-art large language model (LLM), to extract the data-centric rules from the given performance dataset. First, we crafted a prompt by providing metadata of the dataset, different statistical summaries, and a few samples from the given dataset that meet the target constraints. LLM harnessing its own generative power, responds back with a JSON structured response, where each entity contains the following three components with a distinct rule name : (1) Expression: A Python expression to express the rule condition (2) Coverage: How \% of given sample(s) usually follows the rule. (3) Explanation: Brief explanation of the rationale behind this rule and why it's meaningful. However, LLM response can be hallucinated and can generate an invalid rule. To tackle the problem, first we parse the rule and apply the rule on the given samples and validate if the rule is accurately formulated. If any rule is invalid, we opt out of using that rule further. Then we apply those rules to our recommended hypothetical samples and label each sample with a $complience\_score$. We define $complience\_score$ threshold as 0.5; if the sample's $complience\_score$ is below the threshold, we reject those samples from further consideration. 

\begin{equation*}
    complience\_score = \frac{\text{Number of rules the sample follows}}{\text{Number of rules}}
\end{equation*}

We improve the fidelity of the pipeline using the rule-based rejection sampling - (1) it provides an automated way to enforce domain consistency, strengthening the recommended samples not only statistically but also as actionable samples, (2) a brief explanation of each rule can provide inherent structures of the dataset which may be overlooked, (3) this approach is both data-driven and adaptive; first, it doesn't require establishing hard-coded rules by humans and removes manual rejection sampling, secondly we can use this method for different performance datasets without re-engineering. 



\subsection{Evaluation}
The counterfactual and recommendation models in \sysname are evaluated using three complementary methods:
\subsubsection{Uncertainty Quantification:}
In this study, we want to evaluate the uncertainty in predicting the target variable of the recommended samples using a predictive model trained on a real dataset.  This uncertainty quantification analysis serves as a critical tool for evaluating the plausibility and reliability of the generated configurations. The goal of this study is to evaluate how confident a predictive model is able to make accurate predictions on the recommended samples. If the counterfactuals lie in regions of the feature space that are well represented in the training data, the model will exhibit low predictive uncertainty, indicating high trust in its outputs. However, in our generation pipeline, we use a predictive model to infer recommendation samples. So, there is a high risk of potential model bias if we use the same predictive model to evaluate our generated samples. By default, we use XGBoost \cite{chen2016xgboost} as our predictive model in our generation pipeline and RandomForest \cite{biau2016random} for uncertainty quantification. 

\textbf{Outlier Detection}

To evaluate the quality and plausibility of generated recommendations is to use outlier detection. Outlier detection demonstrates a data-centric, distribution-based assessment of how well the generated configurations are aligned with the statistical properties of the real data. The rationale behind using unsupervised outlier detection: generated samples which are not outliers should lie within the learned data manifold of the original distribution. A low number of outliers generated by the outlier algorithms suggests that the generated samples are within the input space trained by the original dataset, which indicates that the generated samples are not only actionable but also realistic. There are different state-of-the-art outlier detection algorithms we can use such as Isolation Forests \cite{liu2008isolation}, One Class SVMs \cite{manevitz2001one} or Local Outlier Factor(LOF) \cite{alghushairy2020review}. We prefer using Isolation Forest(IF) to other density-based and distance-based methods for evaluating  because of its  efficiency in high-dimensional spaces, robustness to feature scaling, ability to detect anomalies without requiring a density or distance threshold, and its suitability for large datasets due to its linear time complexity and low memory footprint compared to other methods.

\textbf{Causal Analysis}

Our evaluation pipeline also includes a causal graph representation of the generated counterfactual samples. The reason behind using causal analysis is to discover the underlying cause-and-effect relationships between different variables of the generated samples. For a generative model, our goal isn't only generating some random samples, but also to provide an understanding of why different variables change in different ways. A causal graph is visualized by a weighted Directed Acyclic Graph (DAG) which encodes directional dependencies among the variables, visual understanding of the interventions, and so on. The edge weight represents the strength of causal influence between connected variables. A high weight indicates the changes of X are likely to produce significant changes in Y, implying a causal relationship.

This evaluation suite provides robustness checks across statistical, predictive, and structural dimensions to ensure the trustworthiness of \sysname's outputs.

\subsection{Explanation}
\label{sec:explanation}
\sysname decouples decision generation from decision interpretation. While Section~\ref{sec:counterfactual} describes the backend that synthesizes candidate configurations, the explanation module interprets these outputs and communicates their implications to stakeholders. The rationale for this separation is that configuration synthesis only answers what is possible; it does not explain 
why it works, how trustworthy it is, and which option best serves the user's goals.

To support informed decision-making, \sysname delivers three forms of explanatory evidence for every generated configuration. First, it provides explanation of what changed and how that impacted the performance metrics using natural language. Second, it offers causal and uncertainty diagnostics to assess trustworthiness. Third, it presents ranked summaries and trade-off metrics to help users choose among feasible alternatives. The next three subsections describe each of these capabilities in detail.

\subsubsection{Interpretability}

For each synthesized configuration \( \mathbf{x}' \), \sysname generates a textual explanation that highlights: (1) what changed from the original \( \mathbf{x} \), (2) the predicted performance impact from \( f(\mathbf{x}') \), and (3) the likely mechanism of improvement (e.g., improved parallelism, memory efficiency). This supports transparency and interpretability by helping users understand the rationale behind each recommendation. To communicate this information,~\sysname presents a table with rows showing the recommendations, predicted metrics, and the generated natural language explanation using any open-source LLM. For our experiments, we used OpenAI's GPT-4o, however, any open source LLM framework such as Ollama~\cite{} provides access to many high-fidelity open-sourced and local models that can be called instead, in the future. 


\subsubsection{Trustworthiness}

To enhance the trustworthiness of the generated samples, \sysname reports two diagnostics for each \( \mathbf{x}' \): (1) causal consistency shown using a violin chart--whether the distributions of the causal influences of each feature matches between the real data and synthesized data, and (2) prediction error and uncertainty showing how confident the model is in its prediction using a MAPE chart with error bars when certain recommended configurations are used. These diagnostics help prevent blind trust in model output and highlight suggestions that are both plausible and stable.


\subsubsection{Decision Support} 

To achieve the ultimate goal of supporting stakeholders make a decision about which suggestion to choose when several recommendations meet the user’s performance goals,~\sysname scores each one using a composite trade-off score, and a radar chart showing how each option balances competing performance metrics.

\section{Experimental Setup}
\subsection{System}

Texas Advanced Computing Center (TACC) provides supercomputing facilities to researchers for conducting simulations. We leverage the Lonestar6 computing cluster for running
all our experiments. Lonestar6 consists of 560 compute nodes
and 88 GPU nodes. Each compute node is comprised of 2
AMD EPYC 7763 64-core (Milan) CPUs and 256 GB of
DDR4 memory; additionally, each of the 84 GPU nodes has
3 NVIDIA A100 GPUs with 40 GB of HBM2 high bandwidth
memory.

\subsection{Graphical User Interface}
We implement a graphical user interface(GUI) using Streamlit to facilitate user interaction with \sysname. We take some user queries from the user interface, and upon generating the configurations, we display two types of information: (1) Top-K recommended configurations in a tabular format, (2) LLM explanation of each of the top-K recommended configurations. As a user, the user will specify the query type defined in \ref{tab:query_templates}.
The interface includes dropdown menus for selecting fixed columns—variables whose values are to remain constant during the analysis—and dropped columns, which are to be excluded from consideration. Additionally, users are required to input JSON objects: (1) A sample dataset which is aligned with the counterfactual what-if, (2) A list of minimum and maximum requirements for the relevant features if needed. Once the target column is selected by the dropdown menu, user will specify the output range for the regression task. After receiving the prior input, our system invokes the generation pipeline to generate feasible, actionable, and plausible configurations. Upon computing a satisfactory score based on the evaluation, the system presents top-K recommendations or hypothetical scenarios in a structured tabular format. To enhance the interpretability, we invoke an \gls{llm} to display a human-readable explanation of the presented tabular format mentioning the changes and their effects. 

\subsection{Datasets}

We use three datasets to run our experiments, which are mentioned in the Table \ref{tab:query_templates}. The PM100 dataset \cite{antici2023pm100} contains 231,116 jobs executed exclusively on the resources of the Marconi100 supercomputer between May and October 2020, with power consumption data recorded at the node, CPU, and memory levels. The Fugaku dataset or F-Data \cite{antici2024f} is a new workload dataset comprising data from approximately 24 million jobs executed on the Fugaku supercomputer during its public usage period from March 2021 to April 2024. We took a subset from those large parquet files. The SC'19 \cite{patki2019performance} dataset introduces a collection of data that explores the trade-offs between optimal and reproducible performance in heterogeneous supercomputing environments by varying multiple system- and user-level parameters simultaneously.

\subsection{Preprocessing}
Our pipeline includes extensive preprocessing. First, we remove the outliers from the dataset. The potential risk of having the outliers is that they skew statistical metrics and reduce the performance of models. Then we remove the zero variance columns, the columns users specified from the dataset. Our next step is to remove the NaN rows to keep our machine learning model robust. We use StandardScaler to transform features to have zero mean and unit variance. We split the dataset into separate train and test sets with common splits of an 80/20 ratio.

\subsection{Generation Pipeline}

We use DiCE \cite{mothilal2020explaining} to implement the backend of our generation pipeline. To extend the diversity in the recommended samples, we leverage an ensemble-based strategy in which we combine multiple DiCE models initialized with different random seeds. However, executing these runs sequentially is computationally expensive. To address this, we integrate MPI-based parallelizm into our generative pipeline. We adopt a static scheduling approach, where each MPI rank is assigned a fixed number of DiCE runs. The rationale behind using static scheduling over dynamic scheduling - we are generating the same $N$ number of counterfactuals using the DiCE generative library which will take roughly the same time for each run. This strategy helps to introduce effective load balancing in our generative pipeline. Upon completion, rank 0 gathers all generated counterfactual samples from the individual ranks and consolidates them into a single csv files for downstream evaluation. 

\subsection{LLM Explanation}

Our end goal is to demonstrate a human-readable explanation of the generated counterfactual. In the graphical user interface, we show the user the top K configurations from the induced configurations; by default, we set $K=5$. To extend our work in the future, we will provide this value configurable by the user. We compute a hybrid approach to determine each sample. To do this, we first sort the configurations based on the distance calculated on the predictive model. This ensures the most optimal solutions to be analyzed first. Once we sort the configurations, we then compute the satisfactory score based on the provided user constraints. We hypothesize that the lower the satisfactory score, the more optimal the outcome, which aligns with one of the key objectives of High Performance Computing: achieving the best performance with minimal resources.

We choose top-k configurations effectively ensuring that we present users with the most optimal and resource-efficient options. We provide the top-k recommendation and the user query to a Large Language Model(LLM) to generate comprehensive, human-readable explanations detailing why each configuration was chosen. Leveraging the robust interpretative capabilities of LLMs, we provide users with clear insights and justification for each recommendation, facilitating informed decision-making. We use OpenAI models to generate the explanation. In the future, we want to provide flexibility to the users to use their own pre-trained LLM model, which will enable seamless integration of an alternative LLM model to provide more customized and robust LLM responses.

\begin{table}[t]
\footnotesize
\centering
\caption{Instantiated Decision-Support Queries Across Datasets and Template Types}
\label{tab:query_templates}
\begin{singlespacing}
\centering
\resizebox{\columnwidth}{!}{
\begin{tabular}{|p{1cm}|p{1.0cm}|p{2cm}|p{2cm}|}
\hline
\textbf{Template Type} & \textbf{Dataset} & \textbf{Instantiated Query} & \textbf{Structured JSON} \\
\hline
Recommendation Query & Fugaku \cite{antici2024f}  & Recommend me a <configuration> where \texttt{duration} is optimal and \texttt{exit state} is $user\_given\_state$ &
\texttt{Type: Recommend} \newline
\texttt{Targets:} \newline
\quad \texttt{- duration: < 1000s} \newline
\texttt{Constraints:} \newline
\quad \texttt{- state}: \texttt{completed}
\\
\hline
Exploratory Optimization Query & PM-100 \cite{antici2023pm100} & Given a base configuration <configuration>, how should I change my current configuration to achieve 'user\_percentage' reduction in \texttt{node\_power\_consumption}? &
\texttt{Type: WhatIf} \newline
\texttt{Baseline: ./inputs/job001.json} \newline
\texttt{Metrics:} \newline
\quad \texttt{- power} \\
\hline
Exploratory What-If Query  & SC'19 \cite{patki2019performance} & Given a base configuration <configuration> what if I double the \texttt{task\_count}, what would be the \texttt{runtime}? &
\texttt{Type: Counterfactual} \newline
\texttt{Baseline: ./inputs/job002.json} \newline
\texttt{Target:} \newline
\quad \texttt{- runtime: 0 - max} \newline
\texttt{Constraints:} \newline
\quad \texttt{- num\_tasks: = 128} \\
\hline
\end{tabular}
}
\end{singlespacing}
\end{table}

\section{Results}
\subsection{Recommendation: Recommend me a <configuration> where \texttt{duration} is optimal and \texttt{exit state} is $user\_given\_state$}

\begin{figure*}[!h]
    \centering
    \subfigure[]{\i
    \includegraphics[width=0.28\textwidth]{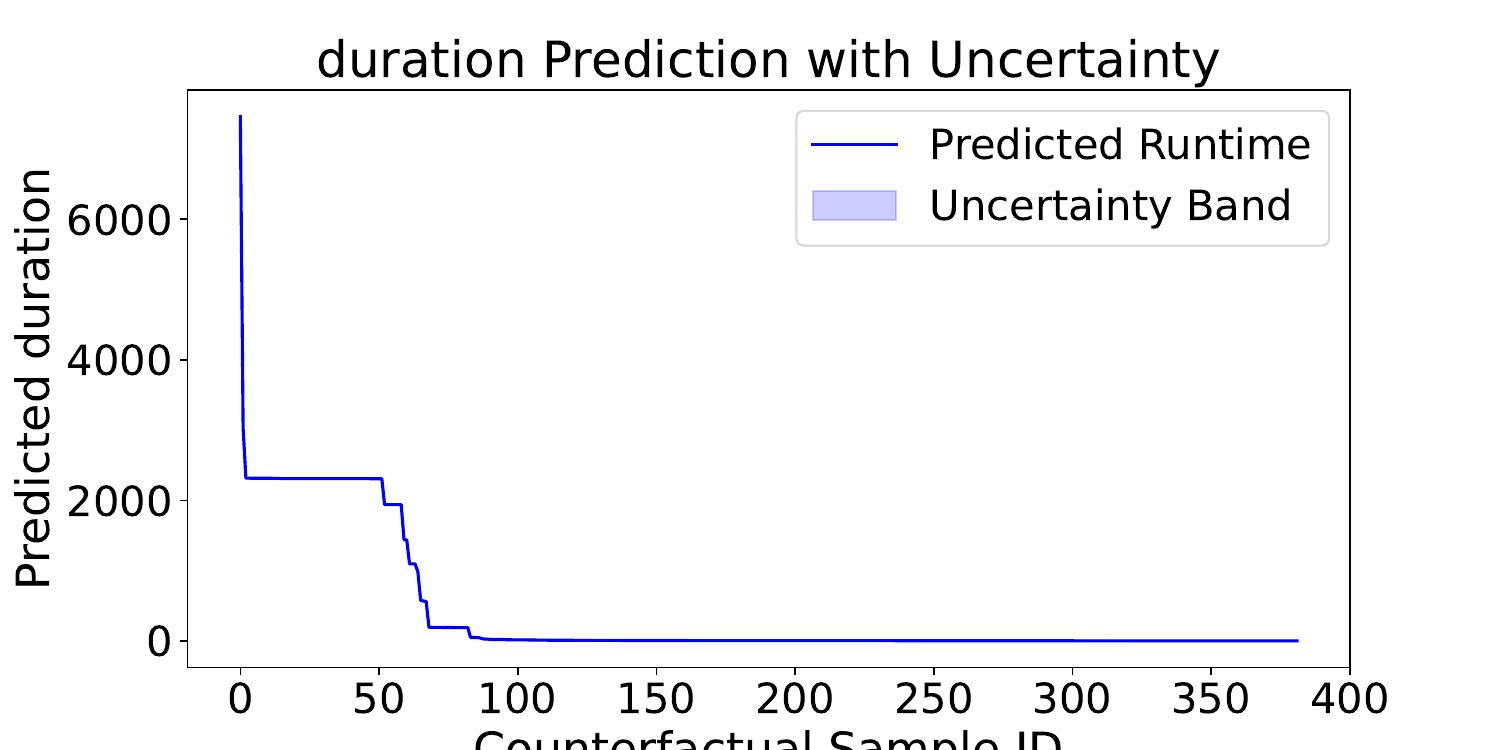}}
    \label{fig:fugaku:uncertainity}
    \hspace{1em}
    \subfigure[]{\includegraphics[width=0.28\textwidth]{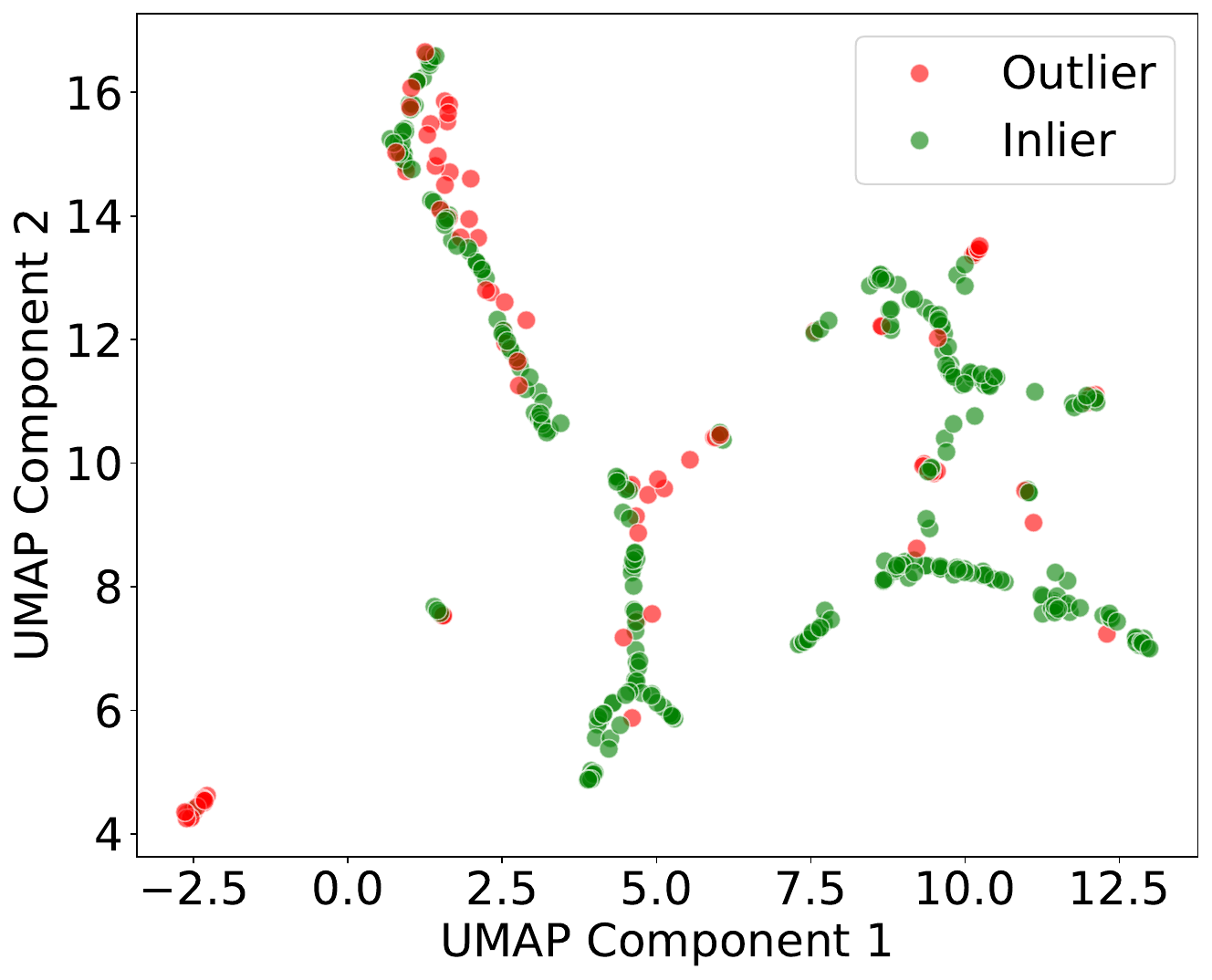}}
    \label{fig:fugaku:outlier}
    \hspace{1em}
    \subfigure[]{\includegraphics[width=0.28\textwidth]{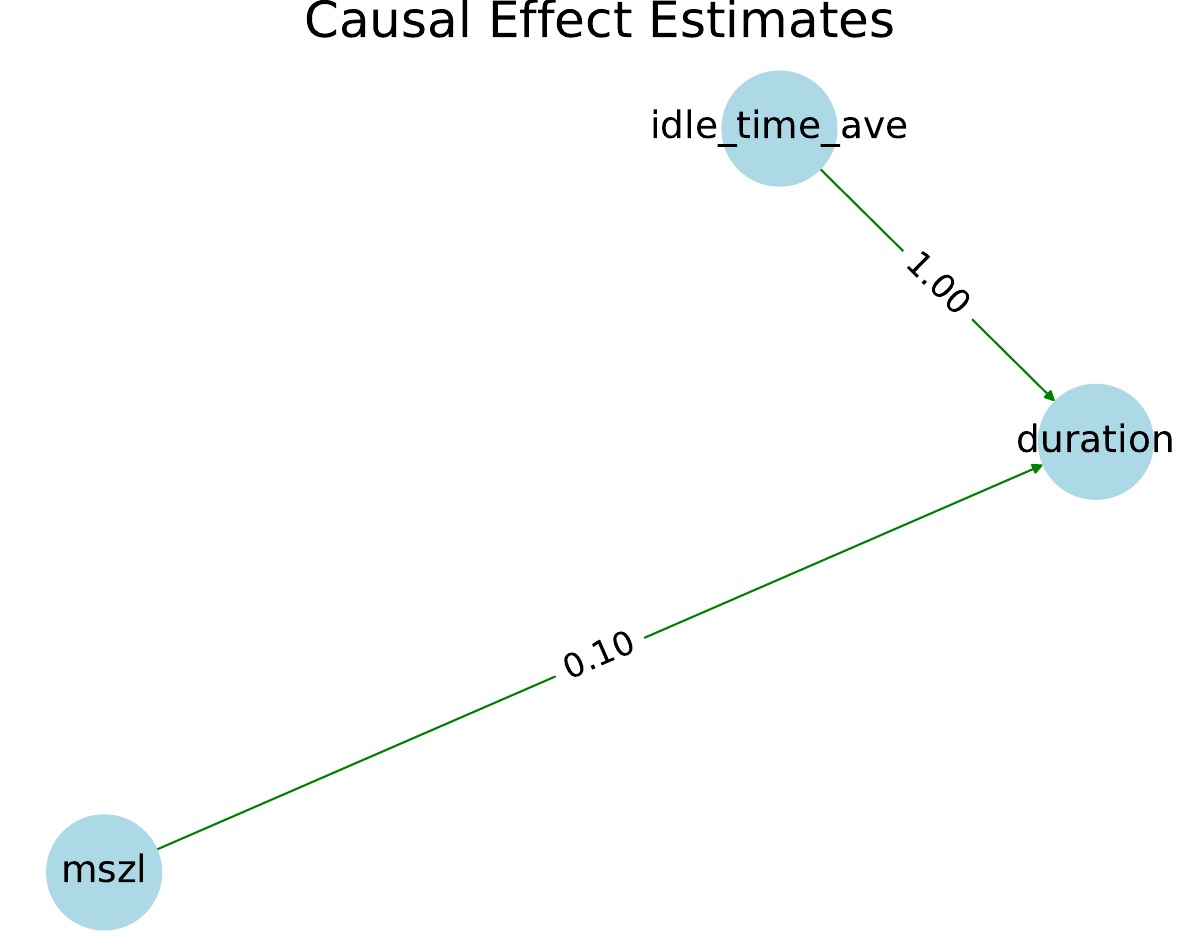}}
    \label{fig:fugaku:causal graph}
    \caption{Results for Fugaku Q1: (a) Uncertainty Quantification, (b) UMAP projection with outlier detection, (c) Causal graph.}
    \label{fig:fugaku}
\end{figure*}

This user query suggests to recommend some hypothetical samples on the Fugaku dataset where the duration is optimal and the exit state strictly follows a user-defined categorical value. Unlike typical numerical constraints where we match the minimum and maximum thresholds, this query introduces column fixing. Our recommendation pipeline produces 382 counterfactual samples, which are aligned with the specified constraints. Our rule-based approach validates all the samples as valid samples. This indicates that 100\% of the samples are valid with respect to user and domain constraints. We analyze the data distribution using our outlier techniques shown in ~\ref{fig:fugaku}, we observe that approximately 22.51\% of the generated samples fall out of the distribution of the training data. This is because of huge variability in the target metrics of the Fugaku dataset. Our query tries to find out the optimal, meaning the engine tries to sample from a lower density distribution.

The uncertainty quantification (UQ) plots demonstrate an exceptionally narrow band. This indicates that the RandomForest model is highly confident in its predictions for the region, which indicates that the model maintains a high level of confidence in its predictions for the generated counterfactuals.

The causal graph indicates that this Fugaku dataset solely depends on the \texttt{idle\_time} metric, and small influenced by the \texttt{mszl}.

\subsection{Exploratory optimization: Given a base configuration <configuration>, how should I change my current configuration to achieve ${user\_percentage}$ reduction in \\$\texttt{node\_power\_consumption}?$}

\begin{figure*}[!h]
     \centering
     \subfigure[]{\includegraphics[width=0.28\textwidth]{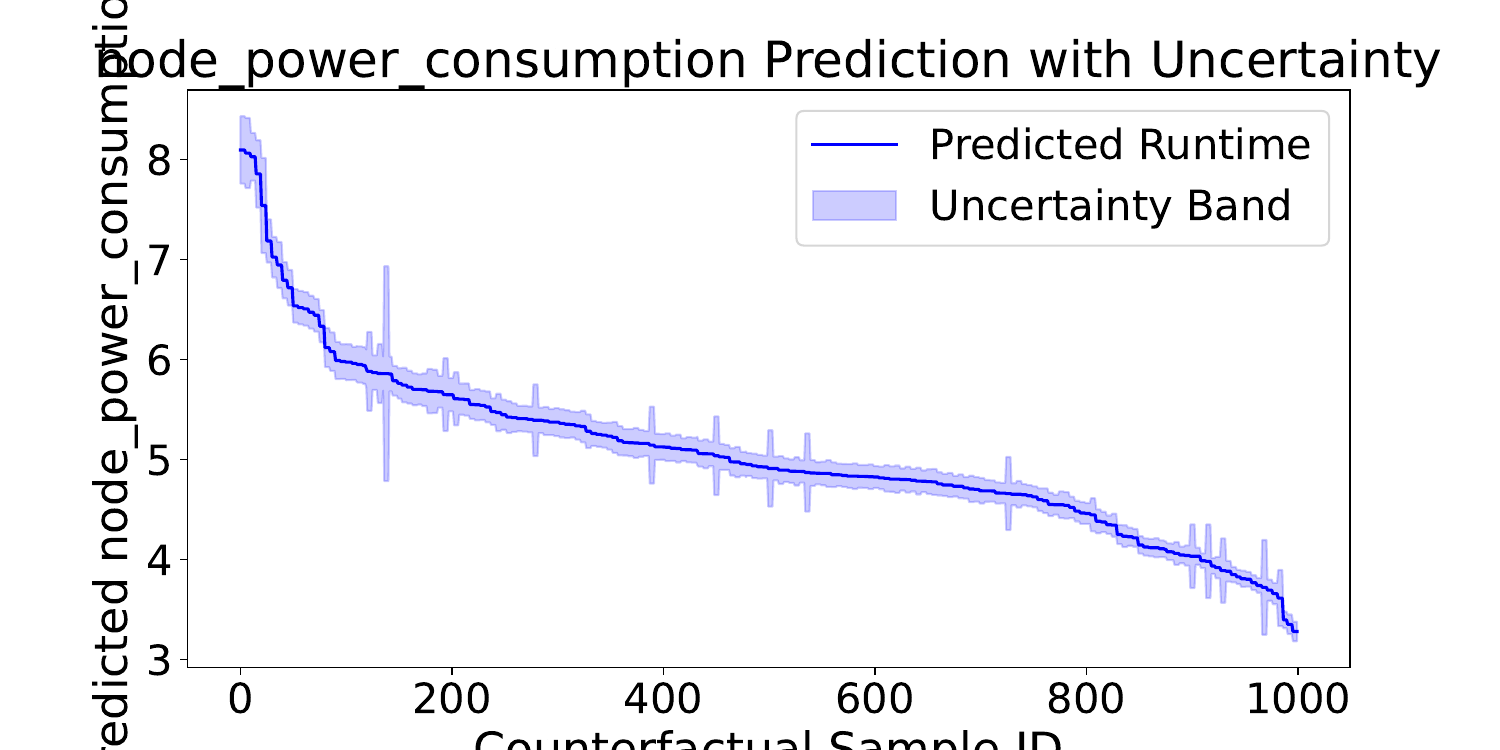}}
     \label{fig:pm100-q2:uq}
     \hspace{1em}
     \subfigure[]{\includegraphics[width=0.28\textwidth]{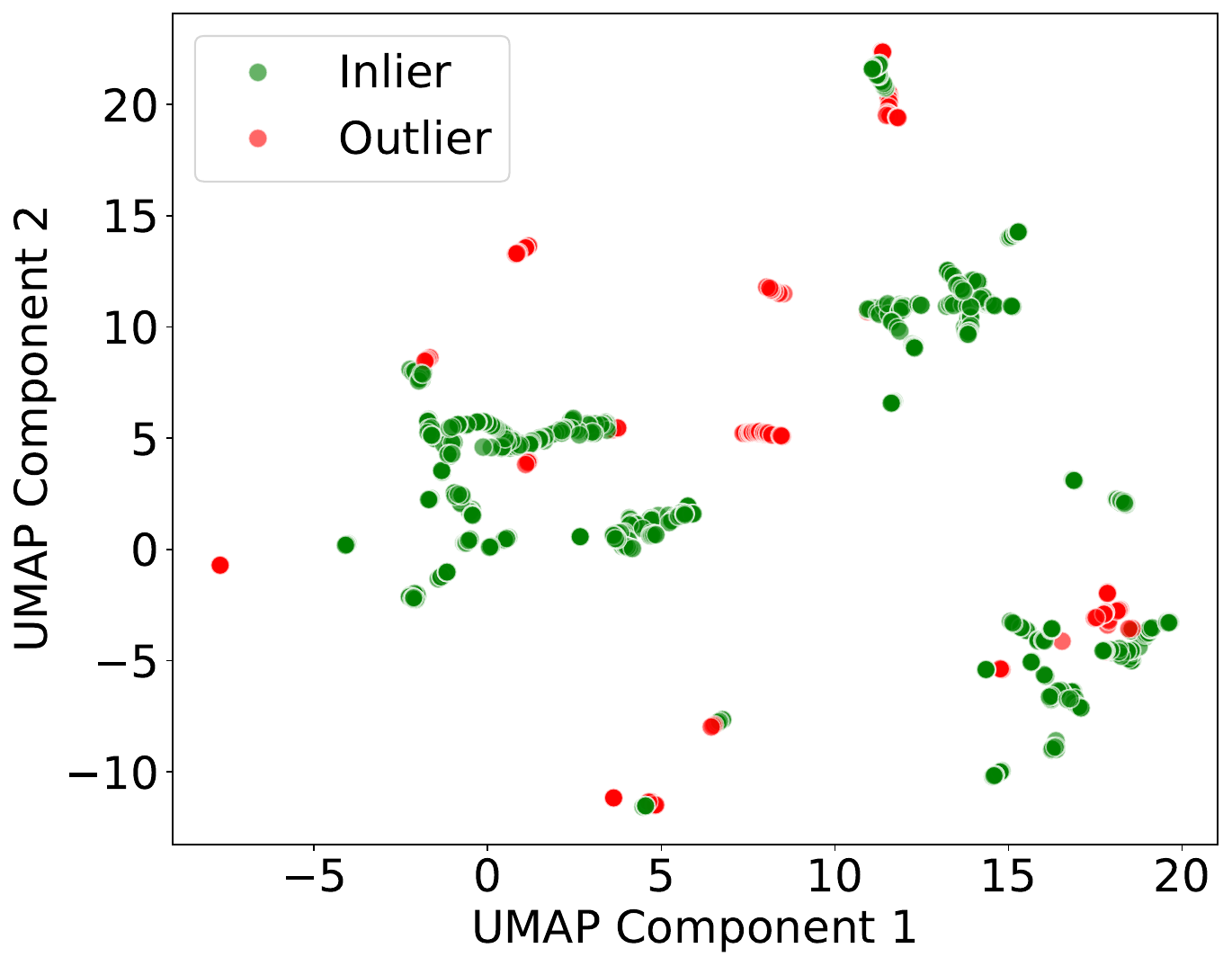}}
     \label{fig:pm100-q2:umap}
     \hspace{1em}
     \subfigure[]{\includegraphics[width=0.28\textwidth]{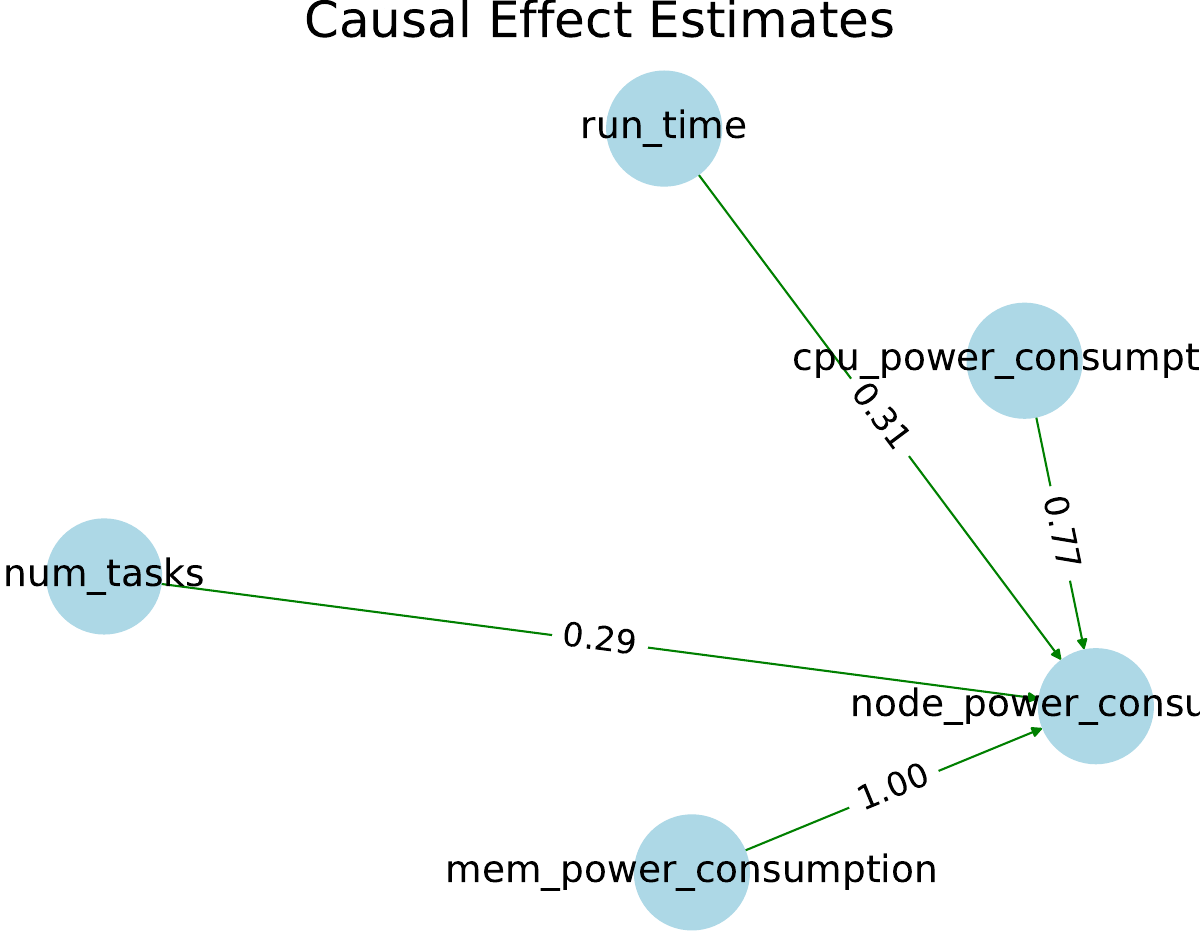}}
    \label{fig:pm100-q2:causal_graph}
     \vspace{-0.5em}
     \caption{Results for PM100-Q2: (a) Uncertainty Quantification, (b) UMAP projection with outlier detection, (c) Causal graph.}
     \label{fig:pm100-q2}
 \end{figure*}

 The primary objective of the query is not to find the best solution as a recommendation from the PM-100 dataset, but also to give users flexibility to explore the configuration space in search of actionable and interpretable configurations. In response to the user query, our generation pipeline offers 1000 alternative samples which could be actionable. Our LLM validation invalidates 8 samples out of 1000 samples. This indicates that our recommendation engine may suggest those samples as valid from their search space; however, they fail to meet the minimum $complience\_score$. 

 Although the rule validation pipeline accepts almost all samples, \ref{fig:pm100-q2} indicates 181 samples out of 1000 samples are outliers. This indicates that a portion of generated data lies in low-density or less represented regions of the original data. This is logical as the rule-based generation method is designed to meet specific constraints or logical conditions — but it may not consider the underlying data distribution. In Figure \ref{fig:pm100-q2}b uncertainty quantification suggests a moderate stability of the model, despite having some random huge spikes. That proves our generated hypothetical samples are prone to the more realistic and sophisticated samples drawn from the training distribution. 

 The causal graph suggests a complex relationship among the features and the target metric. 
 We observe that \texttt{mem\_power\_consumption} and \texttt{cpu\_power\_consumption} are both influential to the
 \\ \texttt{node\_power\_consumption}.

     \begin{figure*}
     \centering
     \subfigure[uq]{
     \includegraphics[width=0.3\textwidth]{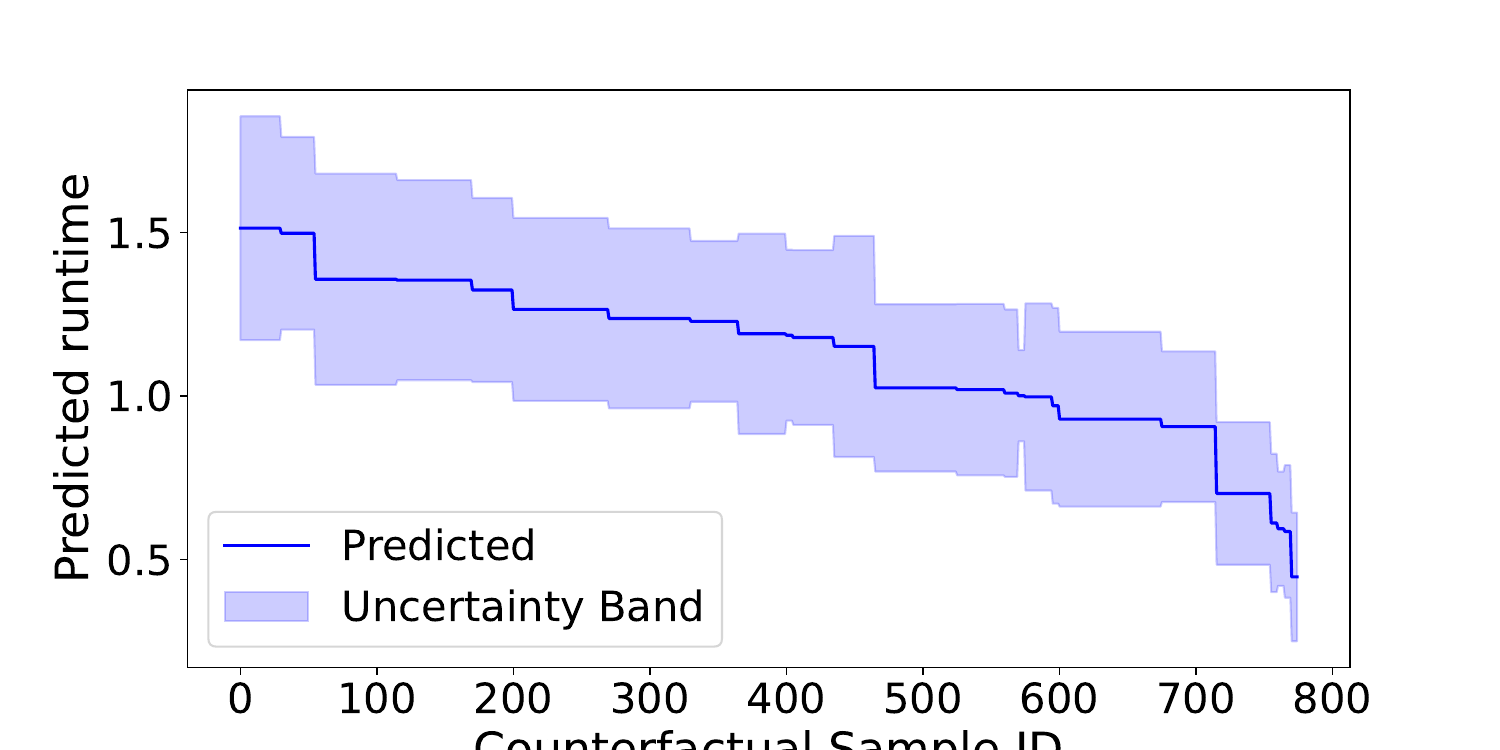}
    \label{fig:UQ-perfvar-q1}
     }
    \subfigure[UMAP]{
     \includegraphics[width=0.3\textwidth]{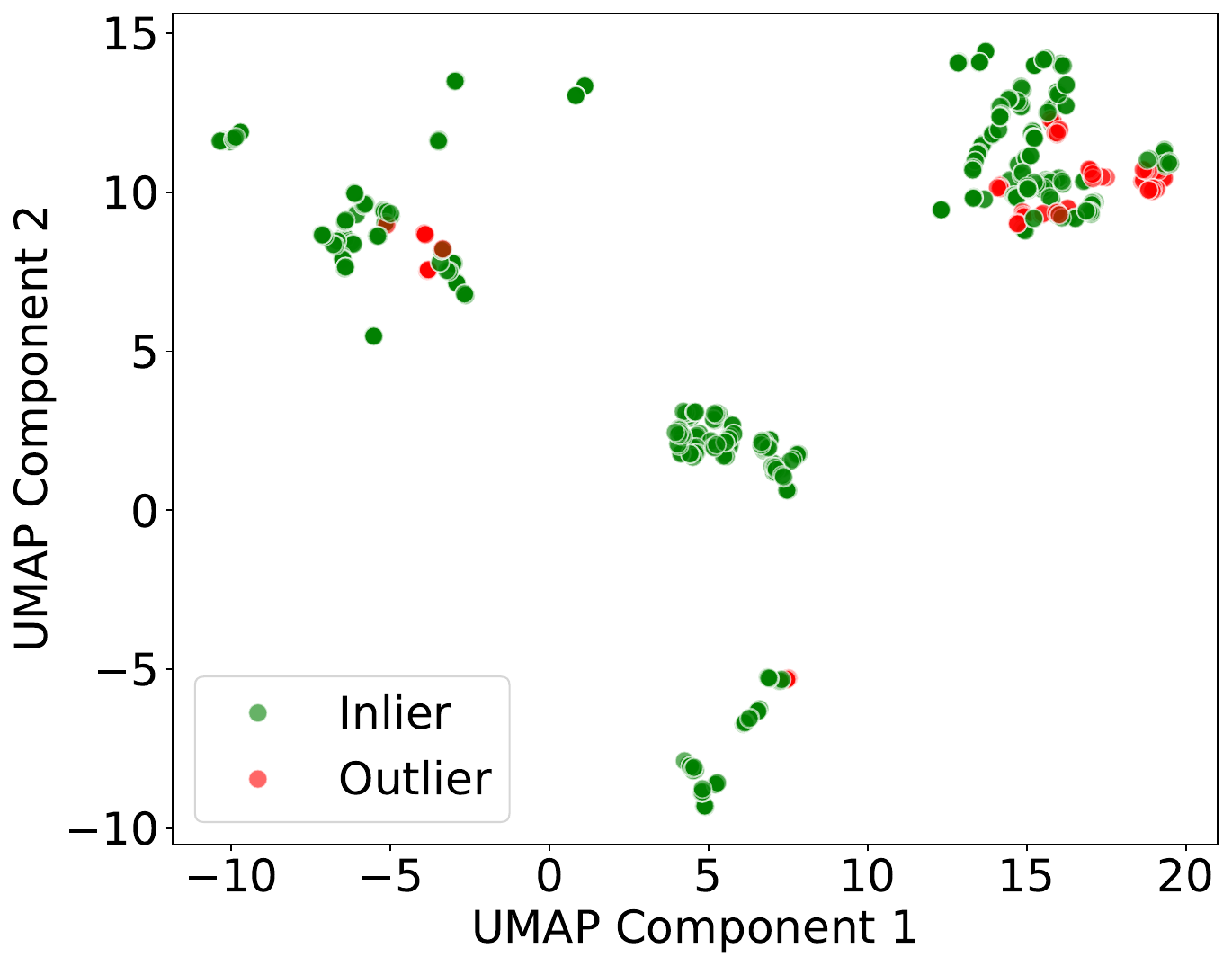}
     \label{fig:UQ-perfvar-q1b}    
     }
     \subfigure[Causal graph]{
     \includegraphics[width=0.3\textwidth]{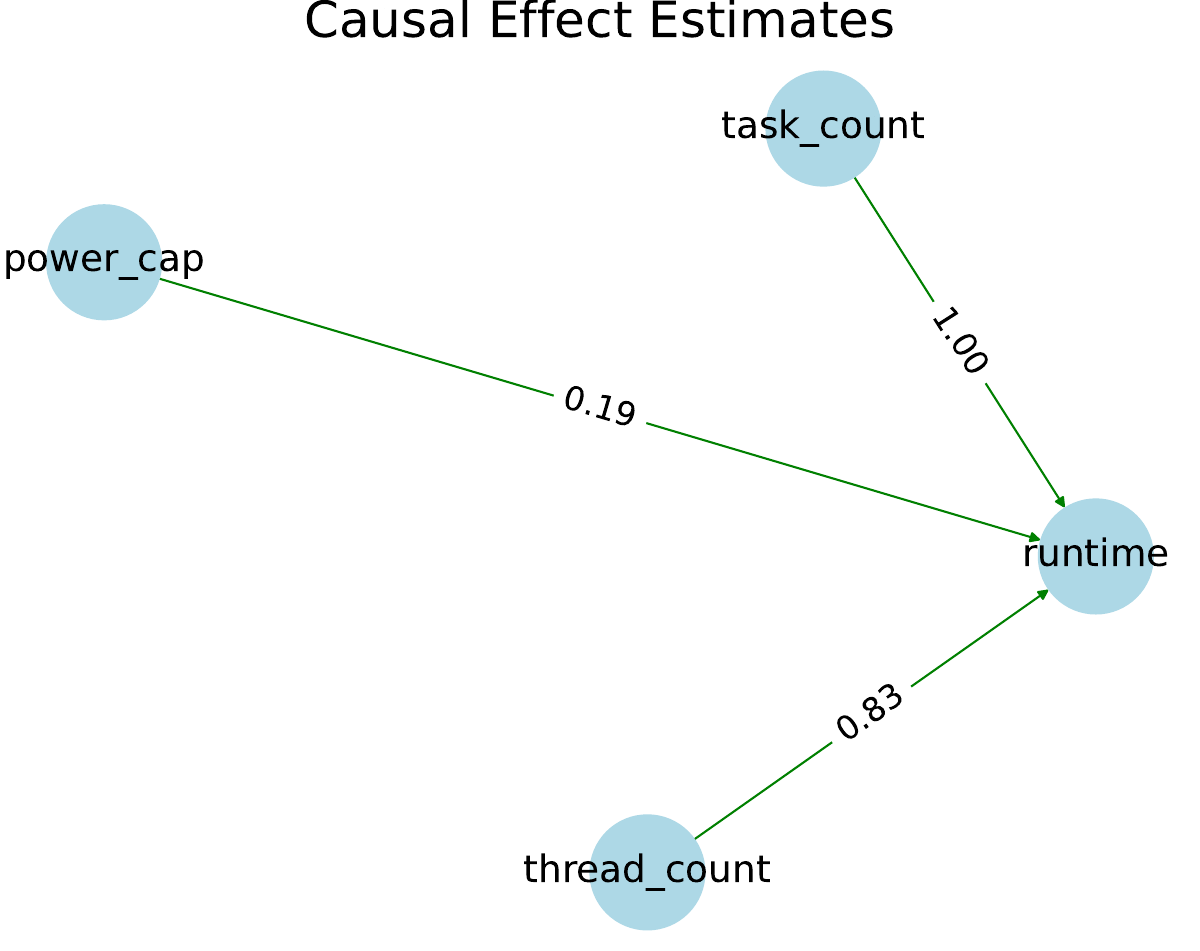}
     \label{fig:UQ-perfvar-q1c}
    }    
     \caption{SC19: (a) Uncertainty Quantification, (b) UMAP projection with outlier detection, (c) Causal graph.}
 \end{figure*}

\subsection{ Exploratory What if: Given a base configuration <configuration> what if I double the \texttt{task\_count}, what would be the \texttt{run\_time?}}

The goal of the query is to investigate the hypothetical scenarios of the SC'19 dataset. The user provides a base configuration and requests the pipeline to explore "what-if" scenarios. Upon processing the requests, our pipeline suggests 775 distinct hypothetical samples. Out of the 775 samples, our rule-based validation mechanism disapproved 16 samples. These samples failed to meet the minimum $complience\_score$ meaning the samples violate more than 50\% of the rules. Figure ~\ref{fig:UQ-perfvar-q1} shows the results of the uncertainty quantification. We see around 91.10\% of samples were confirmed to be valid and actionable. These samples not only align with the data distribution learned from the original dataset but also satisfy practical constraints, indicating high plausibility. 

However, the figure shows wide uncertainty bands in the region. This implies the RandomForest predictive model exhibits lower confidence in sample regions. 

The causal graph exactly matches with our exploration. We find out that the user wants to explore the effect of  \texttt{task\_count} on runtime. We observe that \texttt{task\_count} has the highest influence on the \texttt{run\_time}, changing {task\_count} must have a potential effect on {run\_time}

\section{Related Work}
Counterfactual explanations play a crucial role in healthcare management by enabling physicians to explore alternative treatment plans and understand their potential outcomes. By providing "what-if" scenarios, they help in assessing the impact of changes in treatments or patient conditions, ensuring informed and transparent decision-making. This fosters better patient care and trust by justifying treatment choices with clear, data-driven reasoning. Wang et al. \cite{wang2021counterfactual} 
propose a counterfactual explanation framework for survival prediction of cardiovascular ICU patients using medical event sequences. Employing a text style-transfer technique, the pipeline generates actionable insights for treatment modifications, improving model interpretability and trust in clinical decision-making. Nagesh \cite{nagesh2023explaining}
argue about existing counterfactual methods providing irrelevant counterfactuals, hence suggesting Counterfactual Variational Autoencoder (CF-VAE) to generate plausible and sparse counterfactual explanations for time-series predictions in healthcare tasks like ICU intervention prediction. Kyrimi 
\cite{kyrimi2025counterfactual}
demonstrated an application of counterfactual reasoning using causal Bayesian networks to healthcare governance by extending the scope of counterfactual reasoning to assess clinical decisions across different stages of care and contextualizing its use within mortality and morbidity review meetings. Prosperi et al. 
\cite{prosperi2020causal}
emphasize the importance of distinguishing prediction from intervention models in healthcare. They propose robust frameworks, including the target trial and transportability, to ensure causal validity in counterfactual predictions derived from observational data. Visualizing Counterfactual Clues on Electrocardiograms (VCCE) 
\cite{tanyel2025interpretable}
enhances the interpretability of machine learning models for myocardial infarction detection using ECGs. By integrating counterfactual explanations, feature engineering, and visualization, it provides actionable, clinician-friendly insights, validated through expert evaluations for improved decision-making. However, the aforementioned works focus on classification tasks that predict discrete categories; our work generates and discusses counterfactuals for given samples in a regression context, emphasizing different methodological challenges and evaluation criteria.

Counterfactual analysis plays a vital role in the financial sector by helping stakeholders explore alternative outcomes of strategic decisions. It enables financial institutions to assess the potential effects of adjustments to investment strategies, lending policies, or market dynamics. By offering insights into hypothetical scenarios, it aids in managing risks, ensuring compliance with regulations, and making data-driven decisions to optimize financial performance and stability.
Gunonu et al. \cite{gunonu2024explainable}
proposed a framework integrating counterfactual explanations with tree-based machine learning models to predict bank failures. They demonstrated that the NICE method, coupled with cost-sensitive approaches, effectively improves model interpretability and handles imbalanced data, aiding actionable risk reduction strategies. Tsirtsis et al. 
\cite{tsirtsis2020decisions}
proposed algorithms for generating optimal decision policies and counterfactual explanations in strategic settings, achieving higher utility and incentivizing self-improvement. Their approach models the problem as a Stackelberg game, where decision-makers provide explanations, and individuals strategically respond. They demonstrated the effectiveness of their solutions in financial contexts, such as loan and credit card decision-making, where the algorithms improved utility for decision-makers while encouraging individuals to enhance their financial status. Nguyen et al. 
\cite{nguyen2024using}
applied SHAP
\cite{lundberg2017unified}%
and DiCE to enhance interpretability and provide actionable insights in financial distress prediction, leveraging a dataset of Vietnamese public companies. Their integration of counterfactual explanations with predictive models like XGBoost and neural networks demonstrates the efficacy of combining accuracy and transparency in addressing financial risks in emerging markets.

\section{Conclusions}
In this work, we developed an ad-hoc counterfactual and recommendation pipeline that processes structured queries and generates hypothetical samples and actionable recommendations. We evaluate our samples using three different ways. Our framework supports interpretability by integrating large language models (LLMs) to produce human-readable explanations of the generated results. While our current implementation focuses solely on regression tasks, this serves as a foundational step toward more comprehensive decision-support systems. In future work, we aim to extend our pipeline to support multi-objective optimization and classification tasks, broadening its applicability across diverse problem settings.

\section{Acknowledgment}
This material is based upon work supported in part by the National Science Foundation under Grant No. 2443561. Also, this material is based upon work supported by the U.S. Department of Energy, Office of Science under Award Number DE-SC0022843. The authors acknowledge the Texas Advanced Computing Center (TACC) at The University of Texas at Austin for providing computational resources that have contributed to the research results reported within this paper. URL: http://www.tacc.utexas.edu.

\bibliographystyle{ACM-Reference-Format}
\bibliography{tzi}


\begin{thebibliography}{30}


\ifx \showCODEN    \undefined \def \showCODEN     #1{\unskip}     \fi
\ifx \showISBNx    \undefined \def \showISBNx     #1{\unskip}     \fi
\ifx \showISBNxiii \undefined \def \showISBNxiii  #1{\unskip}     \fi
\ifx \showISSN     \undefined \def \showISSN      #1{\unskip}     \fi
\ifx \showLCCN     \undefined \def \showLCCN      #1{\unskip}     \fi
\ifx \shownote     \undefined \def \shownote      #1{#1}          \fi
\ifx \showarticletitle \undefined \def \showarticletitle #1{#1}   \fi
\ifx \showURL      \undefined \def \showURL       {\relax}        \fi
\providecommand\bibfield[2]{#2}
\providecommand\bibinfo[2]{#2}
\providecommand\natexlab[1]{#1}
\providecommand\showeprint[2][]{arXiv:#2}

\bibitem[Alghushairy et~al\mbox{.}(2020)]%
        {alghushairy2020review}
\bibfield{author}{\bibinfo{person}{Omar Alghushairy}, \bibinfo{person}{Raed Alsini}, \bibinfo{person}{Terence Soule}, {and} \bibinfo{person}{Xiaogang Ma}.} \bibinfo{year}{2020}\natexlab{}.
\newblock \showarticletitle{A review of local outlier factor algorithms for outlier detection in big data streams}.
\newblock \bibinfo{journal}{\emph{Big Data and Cognitive Computing}} \bibinfo{volume}{5}, \bibinfo{number}{1} (\bibinfo{year}{2020}), \bibinfo{pages}{1}.
\newblock


\bibitem[Amershi et~al\mbox{.}(2019)]%
        {amershi2019guidelines}
\bibfield{author}{\bibinfo{person}{Saleema Amershi}, \bibinfo{person}{Dan Weld}, \bibinfo{person}{Mihaela Vorvoreanu}, \bibinfo{person}{Adam Fourney}, \bibinfo{person}{Besmira Nushi}, \bibinfo{person}{Penny Collisson}, \bibinfo{person}{Jina Suh}, \bibinfo{person}{Shamsi Iqbal}, \bibinfo{person}{Paul~N Bennett}, \bibinfo{person}{Kori Inkpen}, {et~al\mbox{.}}} \bibinfo{year}{2019}\natexlab{}.
\newblock \showarticletitle{Guidelines for human-AI interaction}. In \bibinfo{booktitle}{\emph{Proceedings of the 2019 chi conference on human factors in computing systems}}. \bibinfo{pages}{1--13}.
\newblock


\bibitem[Antici et~al\mbox{.}(2024)]%
        {antici2024f}
\bibfield{author}{\bibinfo{person}{Francesco Antici}, \bibinfo{person}{Andrea Bartolini}, \bibinfo{person}{Jens Domke}, \bibinfo{person}{Zeynep Kiziltan}, \bibinfo{person}{Keiji Yamamoto}, {et~al\mbox{.}}} \bibinfo{year}{2024}\natexlab{}.
\newblock \showarticletitle{F-DATA: A Fugaku Workload Dataset for Job-centric Predictive Modelling in HPC Systems}.
\newblock  (\bibinfo{year}{2024}).
\newblock


\bibitem[Antici et~al\mbox{.}(2023)]%
        {antici2023pm100}
\bibfield{author}{\bibinfo{person}{Francesco Antici}, \bibinfo{person}{Mohsen Seyedkazemi~Ardebili}, \bibinfo{person}{Andrea Bartolini}, {and} \bibinfo{person}{Zeynep Kiziltan}.} \bibinfo{year}{2023}\natexlab{}.
\newblock \showarticletitle{PM100: A job power consumption dataset of a large-scale production HPC system}. In \bibinfo{booktitle}{\emph{Proceedings of the SC'23 Workshops of the International Conference on High Performance Computing, Network, Storage, and Analysis}}. \bibinfo{pages}{1812--1819}.
\newblock


\bibitem[Biau and Scornet(2016)]%
        {biau2016random}
\bibfield{author}{\bibinfo{person}{G{\'e}rard Biau} {and} \bibinfo{person}{Erwan Scornet}.} \bibinfo{year}{2016}\natexlab{}.
\newblock \showarticletitle{A random forest guided tour}.
\newblock \bibinfo{journal}{\emph{Test}} \bibinfo{volume}{25}, \bibinfo{number}{2} (\bibinfo{year}{2016}), \bibinfo{pages}{197--227}.
\newblock


\bibitem[Bongers et~al\mbox{.}(2021)]%
        {bongers2021foundations}
\bibfield{author}{\bibinfo{person}{Stephan Bongers}, \bibinfo{person}{Patrick Forr{\'e}}, \bibinfo{person}{Jonas Peters}, {and} \bibinfo{person}{Joris~M Mooij}.} \bibinfo{year}{2021}\natexlab{}.
\newblock \showarticletitle{Foundations of structural causal models with cycles and latent variables}.
\newblock \bibinfo{journal}{\emph{The Annals of Statistics}} \bibinfo{volume}{49}, \bibinfo{number}{5} (\bibinfo{year}{2021}), \bibinfo{pages}{2885--2915}.
\newblock


\bibitem[Chen et~al\mbox{.}(2020)]%
        {chen2020efficient}
\bibfield{author}{\bibinfo{person}{Chong Chen}, \bibinfo{person}{Min Zhang}, \bibinfo{person}{Yongfeng Zhang}, \bibinfo{person}{Weizhi Ma}, \bibinfo{person}{Yiqun Liu}, {and} \bibinfo{person}{Shaoping Ma}.} \bibinfo{year}{2020}\natexlab{}.
\newblock \showarticletitle{Efficient heterogeneous collaborative filtering without negative sampling for recommendation}. In \bibinfo{booktitle}{\emph{Proceedings of the AAAI conference on artificial intelligence}}, Vol.~\bibinfo{volume}{34}. \bibinfo{pages}{19--26}.
\newblock


\bibitem[Chen and Guestrin(2016)]%
        {chen2016xgboost}
\bibfield{author}{\bibinfo{person}{Tianqi Chen} {and} \bibinfo{person}{Carlos Guestrin}.} \bibinfo{year}{2016}\natexlab{}.
\newblock \showarticletitle{{XGBoost}: A scalable tree boosting system}. In \bibinfo{booktitle}{\emph{Proceedings of the 22nd acm sigkdd international conference on knowledge discovery and data mining}}. \bibinfo{pages}{785--794}.
\newblock


\bibitem[Chen et~al\mbox{.}(2015)]%
        {chen2015xgboost}
\bibfield{author}{\bibinfo{person}{Tianqi Chen}, \bibinfo{person}{Tong He}, \bibinfo{person}{Michael Benesty}, \bibinfo{person}{Vadim Khotilovich}, \bibinfo{person}{Yuan Tang}, \bibinfo{person}{Hyunsu Cho}, \bibinfo{person}{Kailong Chen}, {et~al\mbox{.}}} \bibinfo{year}{2015}\natexlab{}.
\newblock \showarticletitle{Xgboost: extreme gradient boosting}.
\newblock \bibinfo{journal}{\emph{R package version 0.4-2}} \bibinfo{volume}{1}, \bibinfo{number}{4} (\bibinfo{year}{2015}), \bibinfo{pages}{1--4}.
\newblock


\bibitem[Deldjoo et~al\mbox{.}(2024)]%
        {deldjoo2024recommendation}
\bibfield{author}{\bibinfo{person}{Yashar Deldjoo}, \bibinfo{person}{Zhankui He}, \bibinfo{person}{Julian McAuley}, \bibinfo{person}{Anton Korikov}, \bibinfo{person}{Scott Sanner}, \bibinfo{person}{Arnau Ramisa}, \bibinfo{person}{Rene Vidal}, \bibinfo{person}{Maheswaran Sathiamoorthy}, \bibinfo{person}{Atoosa Kasrizadeh}, \bibinfo{person}{Silvia Milano}, {et~al\mbox{.}}} \bibinfo{year}{2024}\natexlab{}.
\newblock \showarticletitle{Recommendation with generative models}.
\newblock \bibinfo{journal}{\emph{arXiv preprint arXiv:2409.15173}} (\bibinfo{year}{2024}).
\newblock


\bibitem[Doshi-Velez and Kim(2017)]%
        {doshi2017towards}
\bibfield{author}{\bibinfo{person}{Finale Doshi-Velez} {and} \bibinfo{person}{Been Kim}.} \bibinfo{year}{2017}\natexlab{}.
\newblock \showarticletitle{Towards A Rigorous Science of Interpretable Machine Learning}.
\newblock \bibinfo{journal}{\emph{arXiv: Machine Learning}} (\bibinfo{year}{2017}).
\newblock
\urldef\tempurl%
\url{https://api.semanticscholar.org/CorpusID:11319376}
\showURL{%
\tempurl}


\bibitem[Golaz et~al\mbox{.}(2019)]%
        {}
\bibfield{author}{\bibinfo{person}{Jean-Christophe Golaz}, \bibinfo{person}{Peter~M Caldwell}, \bibinfo{person}{Luke~P Van~Roekel}, \bibinfo{person}{Mark~R Petersen}, \bibinfo{person}{Qi Tang}, \bibinfo{person}{Jonathan~D Wolfe}, \bibinfo{person}{Guta Abeshu}, \bibinfo{person}{Valentine Anantharaj}, \bibinfo{person}{Xylar~S Asay-Davis}, \bibinfo{person}{David~C Bader}, {et~al\mbox{.}}} \bibinfo{year}{2019}\natexlab{}.
\newblock \showarticletitle{The DOE E3SM coupled model version 1: Overview and evaluation at standard resolution}.
\newblock \bibinfo{journal}{\emph{Journal of Advances in Modeling Earth Systems}} \bibinfo{volume}{11}, \bibinfo{number}{7} (\bibinfo{year}{2019}), \bibinfo{pages}{2089--2129}.
\newblock


\bibitem[Gunonu et~al\mbox{.}(2024)]%
        {gunonu2024explainable}
\bibfield{author}{\bibinfo{person}{Seyma Gunonu}, \bibinfo{person}{Gizem Altun}, {and} \bibinfo{person}{Mustafa Cavus}.} \bibinfo{year}{2024}\natexlab{}.
\newblock \showarticletitle{Explainable bank failure prediction models: Counterfactual explanations to reduce the failure risk}.
\newblock \bibinfo{journal}{\emph{arXiv preprint arXiv:2407.11089}} (\bibinfo{year}{2024}).
\newblock


\bibitem[Koren et~al\mbox{.}(2009)]%
        {koren2009matrix}
\bibfield{author}{\bibinfo{person}{Yehuda Koren}, \bibinfo{person}{Robert Bell}, {and} \bibinfo{person}{Chris Volinsky}.} \bibinfo{year}{2009}\natexlab{}.
\newblock \showarticletitle{Matrix factorization techniques for recommender systems}.
\newblock \bibinfo{journal}{\emph{Computer}} \bibinfo{volume}{42}, \bibinfo{number}{8} (\bibinfo{year}{2009}), \bibinfo{pages}{30--37}.
\newblock


\bibitem[Kules and Capra(2009)]%
        {kules2008designing}
\bibfield{author}{\bibinfo{person}{Bill Kules} {and} \bibinfo{person}{Robert Capra}.} \bibinfo{year}{2009}\natexlab{}.
\newblock \showarticletitle{Designing exploratory search tasks for user studies of information seeking support systems}. In \bibinfo{booktitle}{\emph{Proceedings of the 9th ACM/IEEE-CS Joint Conference on Digital Libraries}} (Austin, TX, USA) \emph{(\bibinfo{series}{JCDL '09})}. \bibinfo{publisher}{Association for Computing Machinery}, \bibinfo{address}{New York, NY, USA}, \bibinfo{pages}{419–420}.
\newblock
\showISBNx{9781605583228}
\href{https://doi.org/10.1145/1555400.1555492}{doi:\nolinkurl{10.1145/1555400.1555492}}


\bibitem[Kyrimi et~al\mbox{.}(2025)]%
        {kyrimi2025counterfactual}
\bibfield{author}{\bibinfo{person}{Evangelia Kyrimi}, \bibinfo{person}{Somayyeh Mossadegh}, \bibinfo{person}{Jared~M Wohlgemut}, \bibinfo{person}{Rebecca~S Stoner}, \bibinfo{person}{Nigel~RM Tai}, {and} \bibinfo{person}{William Marsh}.} \bibinfo{year}{2025}\natexlab{}.
\newblock \showarticletitle{Counterfactual reasoning using causal Bayesian networks as a healthcare governance tool}.
\newblock \bibinfo{journal}{\emph{International journal of medical informatics}}  \bibinfo{volume}{193} (\bibinfo{year}{2025}), \bibinfo{pages}{105681}.
\newblock


\bibitem[Liu et~al\mbox{.}(2008)]%
        {liu2008isolation}
\bibfield{author}{\bibinfo{person}{Fei~Tony Liu}, \bibinfo{person}{Kai~Ming Ting}, {and} \bibinfo{person}{Zhi-Hua Zhou}.} \bibinfo{year}{2008}\natexlab{}.
\newblock \showarticletitle{Isolation forest}. In \bibinfo{booktitle}{\emph{2008 eighth ieee international conference on data mining}}. IEEE, \bibinfo{pages}{413--422}.
\newblock


\bibitem[Lundberg and Lee(2017)]%
        {lundberg2017unified}
\bibfield{author}{\bibinfo{person}{Scott~M Lundberg} {and} \bibinfo{person}{Su-In Lee}.} \bibinfo{year}{2017}\natexlab{}.
\newblock \showarticletitle{A unified approach to interpreting model predictions}. In \bibinfo{booktitle}{\emph{Proceedings of the 31st international conference on neural information processing systems}}. \bibinfo{pages}{4768--4777}.
\newblock


\bibitem[Manevitz and Yousef(2001)]%
        {manevitz2001one}
\bibfield{author}{\bibinfo{person}{Larry~M Manevitz} {and} \bibinfo{person}{Malik Yousef}.} \bibinfo{year}{2001}\natexlab{}.
\newblock \showarticletitle{One-class SVMs for document classification}.
\newblock \bibinfo{journal}{\emph{Journal of machine Learning research}} \bibinfo{volume}{2}, \bibinfo{number}{Dec} (\bibinfo{year}{2001}), \bibinfo{pages}{139--154}.
\newblock


\bibitem[Mothilal et~al\mbox{.}(2020)]%
        {mothilal2020explaining}
\bibfield{author}{\bibinfo{person}{Ramaravind~K Mothilal}, \bibinfo{person}{Amit Sharma}, {and} \bibinfo{person}{Chenhao Tan}.} \bibinfo{year}{2020}\natexlab{}.
\newblock \showarticletitle{Explaining machine learning classifiers through diverse counterfactual explanations}. In \bibinfo{booktitle}{\emph{Proceedings of the 2020 conference on fairness, accountability, and transparency}}. \bibinfo{pages}{607--617}.
\newblock


\bibitem[Nagesh et~al\mbox{.}(2023)]%
        {nagesh2023explaining}
\bibfield{author}{\bibinfo{person}{Supriya Nagesh}, \bibinfo{person}{Nina Mishra}, \bibinfo{person}{Yonatan Naamad}, \bibinfo{person}{James~M Rehg}, \bibinfo{person}{Mehul~A Shah}, {and} \bibinfo{person}{Alexei Wagner}.} \bibinfo{year}{2023}\natexlab{}.
\newblock \showarticletitle{Explaining a machine learning decision to physicians via counterfactuals}. In \bibinfo{booktitle}{\emph{Conference on Health, Inference, and Learning}}. PMLR, \bibinfo{pages}{556--577}.
\newblock


\bibitem[Nguyen et~al\mbox{.}(2024)]%
        {nguyen2024using}
\bibfield{author}{\bibinfo{person}{Minh Nguyen}, \bibinfo{person}{Thanh Ngo}, \bibinfo{person}{Bang Nguyen}, {and} \bibinfo{person}{Sukhwa Hong}.} \bibinfo{year}{2024}\natexlab{}.
\newblock \showarticletitle{Using Machine Learning and Counterfactual Explanations for Financial Distress Prediction}.
\newblock \bibinfo{journal}{\emph{Available at SSRN 5032226}} (\bibinfo{year}{2024}).
\newblock


\bibitem[Patki et~al\mbox{.}(2019)]%
        {patki2019performance}
\bibfield{author}{\bibinfo{person}{Tapasya Patki}, \bibinfo{person}{Jayaraman~J Thiagarajan}, \bibinfo{person}{Alexis Ayala}, {and} \bibinfo{person}{Tanzima~Z Islam}.} \bibinfo{year}{2019}\natexlab{}.
\newblock \showarticletitle{Performance optimality or reproducibility: that is the question}. In \bibinfo{booktitle}{\emph{International Conference for High Performance Computing, Networking, Storage and Analysis}}. \bibinfo{pages}{1--30}.
\newblock


\bibitem[Prosperi et~al\mbox{.}(2020)]%
        {prosperi2020causal}
\bibfield{author}{\bibinfo{person}{Mattia Prosperi}, \bibinfo{person}{Yi Guo}, \bibinfo{person}{Matt Sperrin}, \bibinfo{person}{James~S Koopman}, \bibinfo{person}{Jae~S Min}, \bibinfo{person}{Xing He}, \bibinfo{person}{Shannan Rich}, \bibinfo{person}{Mo Wang}, \bibinfo{person}{Iain~E Buchan}, {and} \bibinfo{person}{Jiang Bian}.} \bibinfo{year}{2020}\natexlab{}.
\newblock \showarticletitle{Causal inference and counterfactual prediction in machine learning for actionable healthcare}.
\newblock \bibinfo{journal}{\emph{Nature Machine Intelligence}} \bibinfo{volume}{2}, \bibinfo{number}{7} (\bibinfo{year}{2020}), \bibinfo{pages}{369--375}.
\newblock


\bibitem[Schafer et~al\mbox{.}(2007)]%
        {schafer2007collaborative}
\bibfield{author}{\bibinfo{person}{J~Ben Schafer}, \bibinfo{person}{Dan Frankowski}, \bibinfo{person}{Jon Herlocker}, {and} \bibinfo{person}{Shilad Sen}.} \bibinfo{year}{2007}\natexlab{}.
\newblock \showarticletitle{Collaborative filtering recommender systems}.
\newblock In \bibinfo{booktitle}{\emph{The adaptive web: methods and strategies of web personalization}}. \bibinfo{publisher}{Springer}, \bibinfo{pages}{291--324}.
\newblock


\bibitem[Tanyel et~al\mbox{.}(2025)]%
        {tanyel2025interpretable}
\bibfield{author}{\bibinfo{person}{Toygar Tanyel}, \bibinfo{person}{Sezgin Atmaca}, \bibinfo{person}{Kaan G{\"o}k{\c{c}}e}, \bibinfo{person}{M~Yi{\u{g}}it Bal{\i}k}, \bibinfo{person}{Arda G{\"u}ler}, \bibinfo{person}{Emre Aslanger}, {and} \bibinfo{person}{{\.I}lkay {\"O}ks{\"u}z}.} \bibinfo{year}{2025}\natexlab{}.
\newblock \showarticletitle{Interpretable ECG analysis for myocardial infarction detection through counterfactuals}.
\newblock \bibinfo{journal}{\emph{Biomedical Signal Processing and Control}}  \bibinfo{volume}{102} (\bibinfo{year}{2025}), \bibinfo{pages}{107227}.
\newblock


\bibitem[Tsirtsis and Gomez~Rodriguez(2020)]%
        {tsirtsis2020decisions}
\bibfield{author}{\bibinfo{person}{Stratis Tsirtsis} {and} \bibinfo{person}{Manuel Gomez~Rodriguez}.} \bibinfo{year}{2020}\natexlab{}.
\newblock \showarticletitle{Decisions, counterfactual explanations and strategic behavior}.
\newblock \bibinfo{journal}{\emph{Advances in Neural Information Processing Systems}}  \bibinfo{volume}{33} (\bibinfo{year}{2020}), \bibinfo{pages}{16749--16760}.
\newblock


\bibitem[Wachter et~al\mbox{.}(2017)]%
        {wachter2017counterfactual}
\bibfield{author}{\bibinfo{person}{Sandra Wachter}, \bibinfo{person}{Brent Mittelstadt}, {and} \bibinfo{person}{Chris Russell}.} \bibinfo{year}{2017}\natexlab{}.
\newblock \showarticletitle{Counterfactual explanations without opening the black box: Automated decisions and the GDPR}.
\newblock \bibinfo{journal}{\emph{Harv. JL \& Tech.}}  \bibinfo{volume}{31} (\bibinfo{year}{2017}), \bibinfo{pages}{841}.
\newblock


\bibitem[Wang et~al\mbox{.}(2021)]%
        {wang2021counterfactual}
\bibfield{author}{\bibinfo{person}{Zhendong Wang}, \bibinfo{person}{Isak Samsten}, {and} \bibinfo{person}{Panagiotis Papapetrou}.} \bibinfo{year}{2021}\natexlab{}.
\newblock \showarticletitle{Counterfactual explanations for survival prediction of cardiovascular ICU patients}. In \bibinfo{booktitle}{\emph{Artificial Intelligence in Medicine: 19th International Conference on Artificial Intelligence in Medicine, AIME 2021, Virtual Event, June 15--18, 2021, Proceedings}}. Springer, \bibinfo{pages}{338--348}.
\newblock


\bibitem[Yu et~al\mbox{.}(2023)]%
        {yu2023self}
\bibfield{author}{\bibinfo{person}{Junliang Yu}, \bibinfo{person}{Hongzhi Yin}, \bibinfo{person}{Xin Xia}, \bibinfo{person}{Tong Chen}, \bibinfo{person}{Jundong Li}, {and} \bibinfo{person}{Zi Huang}.} \bibinfo{year}{2023}\natexlab{}.
\newblock \showarticletitle{Self-supervised learning for recommender systems: A survey}.
\newblock \bibinfo{journal}{\emph{IEEE Transactions on Knowledge and Data Engineering}} \bibinfo{volume}{36}, \bibinfo{number}{1} (\bibinfo{year}{2023}), \bibinfo{pages}{335--355}.
\newblock


\end{thebibliography}

\end{document}